# Green and Sustainable Hydrogen-Anchored Solvent Enabling Stable Aqueous Zn Batteries

I. Al Kathemi[a], J. Caroni[b], T. Dehne[a], M. Souto[b,c], M. Antonietti[a], and R. Bouchal[a*]

[a] *Department of Colloid Chemistry, Max-Planck Institute of Colloids and Interfaces, Am Mühlenberg 1, 14476 Potsdam, Germany*

[b] *Centro Singular de Investigación en Química Biolóxica e Materiais Moleculares Departamento de Química-Física Universidade de Santiago de Compostela, Santiago de Compostela 15782, Spain.*

[c] *Oportunius, Galician Innovation Agency (GAIN), Santiago de Compostela, 15702 Spain*

corresponding author: roza.bouchal@mpikg.mpg.de

## Abstract

Aqueous zinc (Zn) batteries provide many benefits, including high theoretical capacity, a low redox potential, and the abundance of Zn in Earth's crust. However, the benefits are often compromised by severe side reactions and dendrite growth, limiting their practical application. To mitigate these drawbacks, many studies have focused on high-concentration electrolytes to desolvate the $Zn^{2+}$ solvation shell from water and to reduce the amount of free water. In this study, a methodology is proposed to achieve the benefits of highly concentrated electrolytes while using low salt concentrations. 5-methyl-2-pyrrolidone (5MP) was introduced as a safe and polar cosolvent capable of anchoring free water molecules through hydrogen bonding, enabled by its carbonyl and secondary amine groups. Additionally, by adding a co-salt containing a chaotropic ion, it is possible to disrupt the water network, enabling the development of high-performance aqueous Zn batteries. The synergy between anchoring water molecules and disrupting the overall water network proved to be an effective strategy for enhancing the overall Zn battery performance in a Zn∥Zn cell, with lifetimes exceeding 2000 hours and 1400 hours at 1 C and 5 C, respectively. Analysis of the recovered Zn anode confirmed that the combination of 5MP and chaotropic ions enabled Zn deposition along the energetically favorable (002) plane, with no signs of surface dehydration observed by in situ Raman spectroscopy. Furthermore, long-term stability tests using a carbonyl-rich COF- and NaVO-based cathodes at various current densities further demonstrated the benefits of this approach. This work showcases the universality of a diluted electrolyte in combination with 5MP and chaotropic ions, bridging the gap between laboratory research and real-world applications.

# Introduction

Aqueous rechargeable metal-ion batteries, particularly aqueous Zn batteries (AZBs), have emerged as promising next-generation energy storage systems due to their intrinsic safety, environmental friendliness, cost-effectiveness, and high ionic conductivity.[1,2] AZBs also benefit from the natural abundance of Zn (approximately 300-fold higher than that of lithium), its high theoretical capacity (820 mAh/g), and a lower electrochemical potential (–0.763 V vs. standard hydrogen electrode). In addition, the non-toxicity and simplified cell assembly make Zn anodes highly attractive compared to other metal anodes in aqueous electrolytes.[3–5] Despite these advantages, AZBs still face significant challenges similar to those of LIBs, including cathode dissolution,[6] Zn dendrite growth,[7] interfacial side reactions at the Zn anode,[8] such as hydrogen evolution and corrosion, as well as the limited chemical and electrochemical stability of the electrolyte.[9] The instability stems from hydrated $[Zn(H_2O)_6]^{2+}$ ion complexes and free water molecules, which trigger severe interfacial side reactions.[10] Consequently, these issues lead to low coulombic efficiency (CE) and may compromise battery safety, resulting in swelling or short-circuit failure.[11]

Several approaches to mitigate these challenges have been tested, including surface modification,[12] optimising the structure of Zn hosts,[13] separator design,[14] and electrolyte modification.[15] Among these, electrolyte design is considered a highly advantageous solution due to its ease of modification and feasibility for scaling up to the application scale.[16] One of the most effective strategies to suppress the hydrogen evolution reaction and corrosion is the employment of the 'water-in-salt' electrolyte (WiSE) concept introduced by Suo et al., for LIBs in 2015[17] and by Wang et al., for AZBs in 2018.[18] In WiSE, the mass or volume of the salt outweighs that of water, which results in an anion-containing solvation sheath with little free water.[19] However, this concept is challenged by high viscosity, high costs, and the use of hazardous chemicals such as bis(trifluoromethylsulfonyl)imide (TFSI) and its analogues.[20] Consequently, alternative strategies are needed that preserve the key advantages of WiSEs while alleviating these limitations. One promising approach is to introduce organic co-solvents to modify the $Zn^{2+}$ solvation structure in dilute electrolytes. By partially substituting water, these co-solvents promote anion- and solvent-rich $Zn^{2+}$ solvation and weaken the hydrogen-bonded water network, thereby suppressing water-induced parasitic reactions.[21–25] Achieving WiSE-like properties with low salt concentrations and minimal co-solvent content, however, remains a challenge. For instance, Ming et al. incorporated propylene carbonate (PC) into 1 molar (M = mole of salt/volume of solution) Zn sulfate ($SO_4$) and 1 M Zn triflate ($(OTf)_2$) electrolytes, optimising the system at 50 vol% PC.[26] Zhang et al. systematically screened multiple anions and organic co-solvents to improve coulombic efficiency in AZBs, ultimately selecting a hybrid bi-salt formulation of 2.8 molal (m = mol of salt / 1 kg of solvent) $Zn(OTf)_2$ + 0.7 m Zn acetate ($Ac_2$) + 30 wt% dimethylformamide (DMF).[27] Furthermore, Li et al. demonstrated that a lower cosolvent content (10 vol% N-methyl-2-pyrrolidone, NMP) in 2 M $ZnSO_4$ could effectively disrupt water clusters via the proton-acceptor ability of the organic solvent.[28] However, such organic solvents, in particular NMP, are generally considered toxic and flammable,[29] and the use of fluorinated salts further raises safety concerns.[30] Consequently, there is a clear need to develop non-toxic solvents and chemicals for the practical application of AZBs.

Herein, a safer alternative to NMP, 5-methyl-2-pyrrolidone (5MP),[31] is introduced as a co-solvent for the first time in electrolyte formulations, to the best of our knowledge. Substituting the methyl group in NMP with a hydrogen atom at the N-H site[32] reduces its interactions with biological targets associated with NMP toxicity, particularly reproductive effects.[33] Moreover, 5MP, with two hydrogen-bonding sites, can better disrupt the water network compared to NMP, which is only a hydrogen bond acceptor. Consequently, 5MP is hypothesized to bind free water more effectively and to suppress spontaneous side reactions in aqueous Zn electrolytes. We further investigate the role of chaotropic cations (Chao$^+$) as a co-salt to disrupt water networks in low-concentration electrolytes. Zn acetate (ZnAc) was combined with additional acetate salts containing Chao$^+$ ions,[34–36] including potassium (K),[37] ammonium (NH$_4$),[38] and guanidinium (Gua).[39] The influence of different Chao$^+$ ions in combination with varying vol% of 5MP on AZB performance was systematically examined. Raman spectroscopy and $^1$H NMR revealed clear correlations between water network structure, chaotropic cation strength, and 5MP interactions, confirming that anchoring free water while disrupting bound water is an effective strategy.

These electrolytes were then also applied to electroactive covalent organic framework (COF)-based cathodes,[40] which have emerged as promising organic electrode materials for various metal-ion batteries,[41–44] including AZBs.[45–48] In particular, the COF constructed from pyrene-4,5,9,10-tetraone units (PT-COF) exhibits excellent long-term cycling stability in concentrated electrolytes.[49,50] New solvent systems, however, should always be evaluated for potential interference with cathode stability. Using PT-COF, we demonstrated the performance of the developed electrolytes in a more sustainable AZB system. To further extend our solvent approach, a vanadium-based cathode was also tested as a proof of concept.[40] Overall, this work presents an electrolyte engineering strategy that combines non-toxic salts and green organic co-solvents at low concentrations, enabling effective, diluted electrolytes for high-performance AZBs.

## Results and discussion

**Electrolyte preparation and characterisation**

To improve electrolyte stability and suppress parasitic reactions in Zn batteries, 5MP, a highly polar aprotic solvent, was introduced to alter the Zn$^{2+}$ solvation structure. Unlike NMP, which has one H-bond acceptor (carbonyl group),[51] 5MP has two H-bond acceptors, namely the carbonyl and secondary amine (Fig. 1a), and a H-bridge donor (again the amine). Therefore, it is expected that 5MP disrupts water–water interactions and partially replaces coordinated water in the Zn$^{2+}$ solvation shell, helping to reduce water activity at the Zn interface better than NMP.[28] Furthermore, the impact of three different chaotropic cations (K$^+$, NH$_4^+$, and C(NH$_2$)$_3^+$ (Gua$^+$)) as co-salts, also referred to as structure breakers, was investigated to further reduce hydrogen bonding in water. These cations exhibit different chaotropic effects, following the trend K$^+$ < NH$_4^+$ < Gua$^+$ as illustrated in Fig. 1b.[52] Different solutions were prepared to investigate the effects of increasing 5MP content in water, in 1 m ZnAc electrolyte, and in 1m ZnAc + 1m **Chao$^+$**Ac (Zn**Chao$^+$**Ac) bi-salt electrolytes (Fig. S1).

First, the effect of 5MP was evaluated in pure water using Raman spectroscopy. As shown in Fig. 1c, an increase in 5MP content led to a decrease in the Raman intensity within the water region. The peak at a higher wavenumber (>3400 cm$^{-1}$), corresponding to weaker hydrogen (H)-bonds, decreased more significantly than the peak at lower wavenumbers (<3300 cm$^{-1}$), which corresponds to stronger H-bonds (Fig. S2a).[53] This suggests that strong H-bonds were more present with increasing 5MP content, even though the total water signal decreased. In ZnAc electrolyte, similar behaviour was observed, where a noticeable reduction in water activity appeared from 40 vol% 5MP (Fig. 1d), but with a higher ratio between strong and weak H-bonds. In bi-salt systems (Fig. S2b-d), the suppression of water signals was already evident at 10 - 20 vol% 5MP, indicating that the chaotropic cations affect the water environment, and therefore, less vol% 5MP was required. Additionally, the viscosity measurements of the 5MP in water solutions (Fig. S3), indicate the increase in viscosity from 10 vol% to 20 vol% 5MP (1.2 and 1.7 mPa·s, respectively) was not significant as a further increase in 5MP content, 3.3 mPa·s for 40 vol% 5MP, 6.7

mPa·s for 60 vol% 5MP and 21 mPa·s for pure 5MP. Therefore, from these results, 20 vol% of 5MP was chosen for all electrolytes (Fig. S4 onwards) with the number 20 indicating the vol% 5MP.

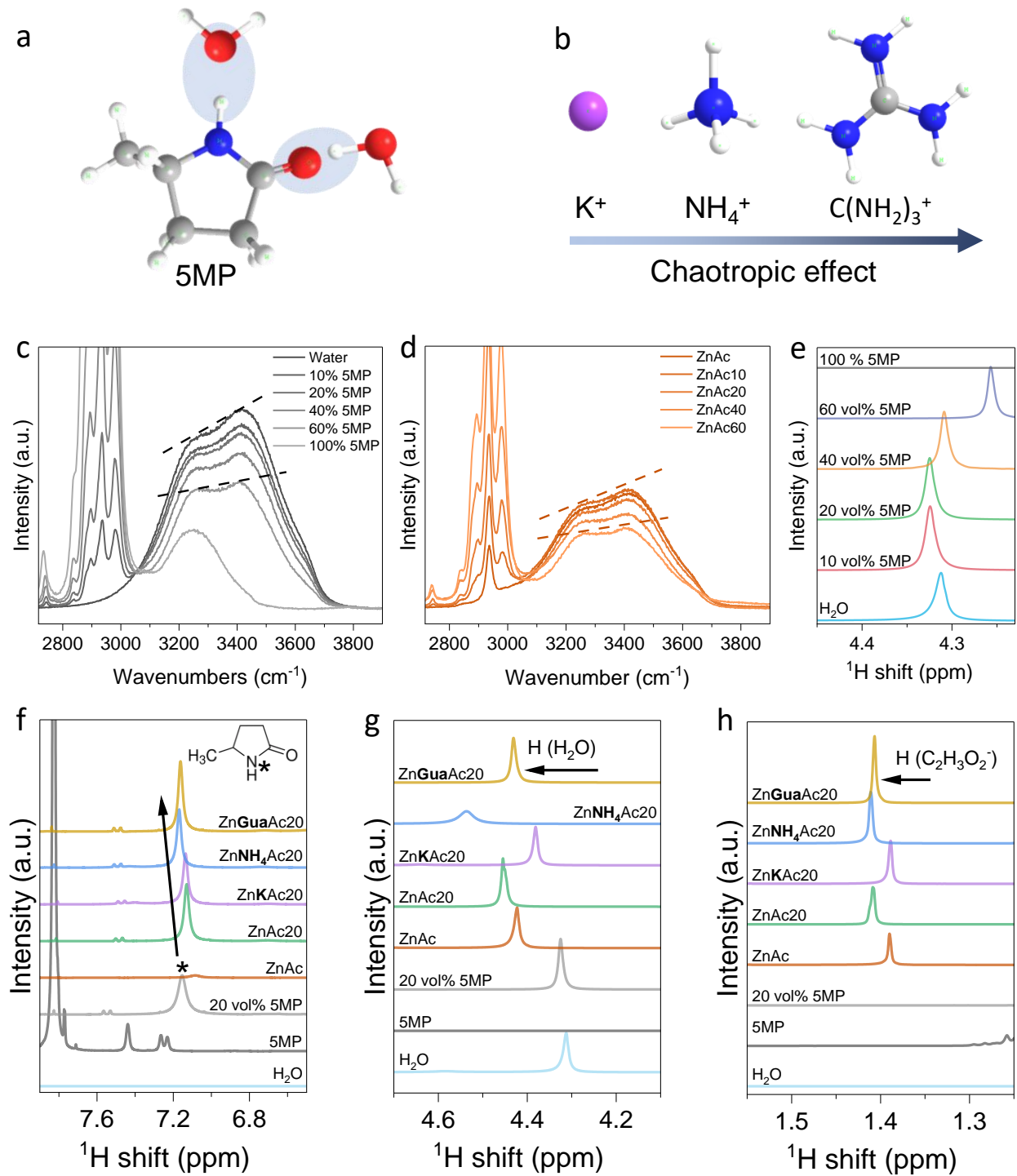

Fig. 1. Characterisation of the electrolytes. a) Molecule structure of 5MP with two H-bonds. b) Molecule structure of $K^+$, $NH_4^+$ and $C(NH_2)_3^+$ ions. c-d) Raman spectra of the water region of pure water and ZnAc with increasing 5MP volume percentage, respectively. e) $^1H$ NMR spectra of water with increasing vol% 5MP. f-h) $^1H$ NMR spectra of the H-bonds of –N-H in 5MP, $H_2O$ and acetate, respectively.

The Raman spectra of the bi-salt electrolytes with 5MP (ZnAc20, Zn**K**Ac20, Zn**NH₄**Ac20, and Zn**Gua**Ac20) revealed that minor shifts in the C-C and C-H₃ vibrations were present (Fig. S4). For a better understanding, the spectra were also compared with single-salt electrolytes of 1 m **K**Ac, 1m **NH₄**Ac, and 1 m **Gua**Ac with 20% 5MP. The C-C stretching at 934 cm$^{-1}$ for acetate ions (Fig. S4a)[54] shifts towards lower wavenumbers (926 cm$^{-1}$) in single-salt electrolytes, indicating a weakened chemical interaction with acetate. However, the C-C stretching vibrations shifted back to a higher wavenumber of 934 cm$^{-1}$ due to the combination of ZnAc with a chaotropic acetate salt, indicating that the coordination of acetate is stronger with the Zn cation. Additionally, the C-H₃ bending of acetate ions shifts to lower wavenumbers (from 1349 cm$^{-1}$ to 1342 cm$^{-1}$ in Fig. S4b)[55] in single- and bi-salt electrolytes, indicating again that acetate has fewer chemical interactions with its surrounding ions. However, the C=O stretching of acetate shifts to higher wavenumbers (from 1418 to 1425 cm$^{-1}$)[54], possibly indicating that the surrounding ions prefer to chemically interact with the C=O of acetate instead of its C-H₃ bond. Unlike the acetate ions, the C-C stretching of 5MP at 954 cm$^{-1}$ [56] did not show any shifts, indicating no significant change in the C-C vibrational modes of 5MP. However, the interaction of 5MP with its surrounding ions still occurred, as indicated by the shift to higher wavenumbers in the C-H₃ bending (from 1339 to 1342 cm$^{-1}$) and stretching (from 1386 to 1391 cm$^{-1}$) of 5MP (Fig. S4b), indicating stronger chemical interactions between 5MP and its surrounding ions in the single- and bi-salt electrolytes.

To further understand the interactions within the electrolytes, proton nuclear magnetic resonance ($^1$H NMR) was conducted using coaxially inserted deuterated dimethyl sulfoxide (DMSO-d₆) as a standard reference solvent. The $^1$H NMR spectra of 5MP in aqueous solutions reveal that increasing water content progressively alters the local solvation environment (Fig. S5). The –CH₃ (Fig. S5a) and N–H (Fig. S5e) protons exhibit upfield shifts at high water fractions (≥90 vol%), consistent with weakened 5MP–5MP interactions as water dominates the solvation shell. At low 5MP content (10–20 vol%), the N–H proton shows a downfield shift, indicative of stronger H-bonding with water, whereas non-labile protons remain less affected. These trends demonstrate that water modulates the balance between intra- and intermolecular interactions in 5MP, progressively replacing solute–solute contacts with water-mediated H-bonds.

In single- and bi-salt electrolytes containing 20 vol% 5MP, non-labile protons generally shift upfield, reflecting weakened interactions with surrounding ions and molecules, including Zn$^{2+}$, chaotropic cations, acetate, and water (Figs. S6a-e). In contrast, the labile N–H proton (Fig. Se) exhibits downfield shifts in the presence of specific cations (e.g., Chao$^+$ or Zn$^{2+}$), indicating stronger H-bonding. The magnitude of these shifts correlates with cation chaotropic strength (Gua$^+$ ≤ NH₄$^+$ < K$^+$), highlighting ion-specific effects on the 5MP solvation environment. Bi-salt electrolytes (Fig. S6f-i) slightly amplify these trends, suggesting cooperative modulation of H-bond networks by multiple cations. Similarly, the water resonance shifts downfield in NH₄$^+$-containing systems, consistent with a strengthened H-bond network, whereas K$^+$-containing electrolytes show minimal change (Fig. S7a-b).

The acetate resonance was analyzed to disentangle the contributions of Zn$^{2+}$ and 5MP. In single-salt electrolytes containing 5MP, acetate chemical shifts are similar to those in bi-salt electrolytes without 5MP (Fig. S7c and d, respectively), suggesting that acetate experiences a comparable H-bonding environment in the presence of either Zn$^{2+}$ or 5MP. Since it is known that the acetate has a preference to bind with Zn$^{2+}$ ions[57], it can be concluded that 5MP has the same binding preference to acetate as Zn$^{2+}$. In bi-salt electrolytes with 20 vol% 5MP (Fig. 1h), acetate shows a clear downfield shift relative to systems containing only 5MP or Zn$^{2+}$ (Fig. S7c and d), supporting the cooperative effect of both chaotropic cations and 5MP in restructuring the H-bond network in low-concentration aqueous electrolytes.

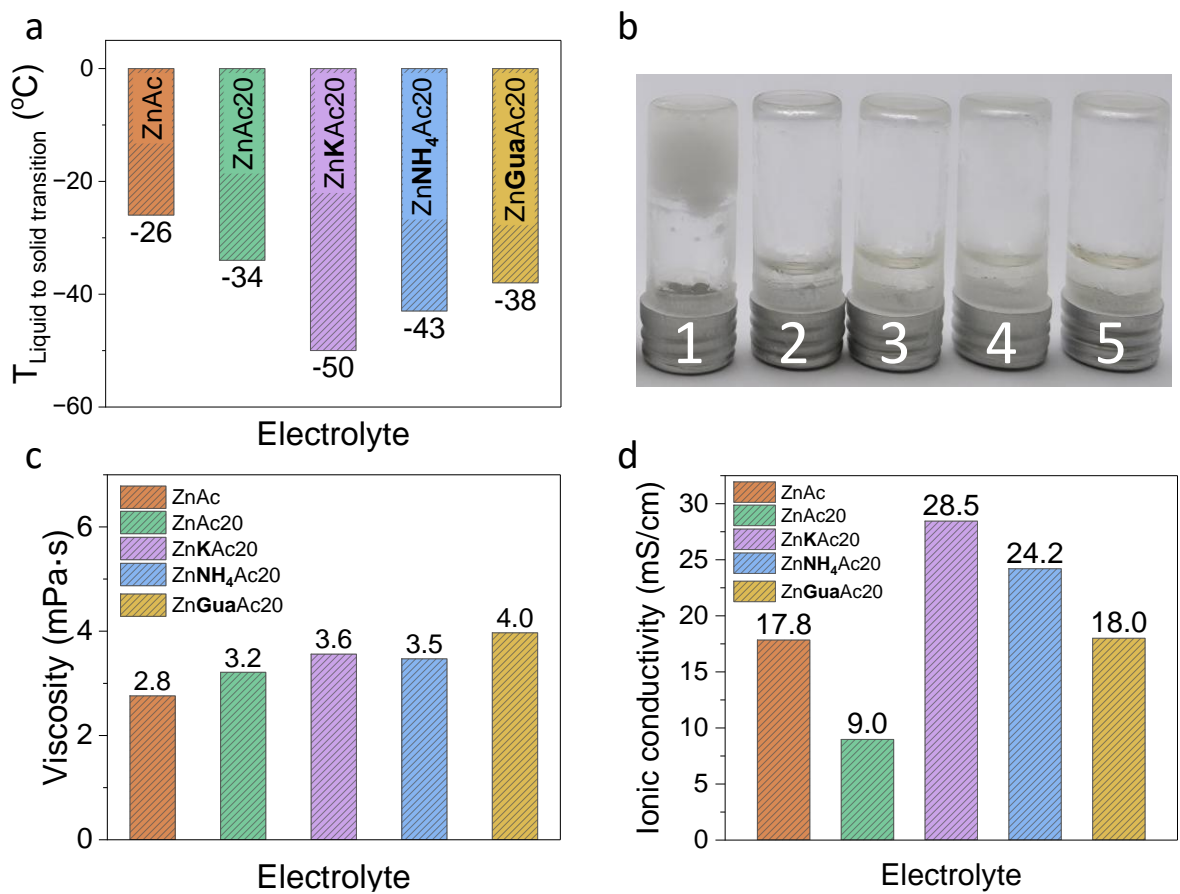

*Fig. 2. Physicochemical properties of the electrolytes. a) Measured crystallisation temperatures. b) Digital images of ZnAc (1), ZnAc20 (2), Zn**K**Ac20 (3), Zn**NH₄**Ac (4), and Zn**Gua**Ac20 (5) after storing at -18 °C for 24 hours. c) Determined viscosity at 25 °C. d) Determined ionic conductivity.*

To test the physicochemical properties, differential scanning calorimetry (DSC) was performed between +50 and -150 °C (Fig. 2a and S8). The results showed that the higher the 5MP volume percentage, the lower the measured transition temperatures, except for pure 5MP, which crystallises at room temperature (Fig. S8a). All bi-salt electrolytes exhibited lower liquid-to-solid (crystallization point, $T_c$) and solid-to-liquid (melting point, $T_c$) transition temperatures than the ZnAc electrolyte, highlighting the water-binding effect of 5MP and the disruption of the water network by chaotropic ions. Notably, both transition temperatures were inversely proportional to the chaotropic strength of the respective ions: Zn**Gua**Ac20, containing the strongest chaotropic ion Gua⁺, showed the highest transition temperatures of $T_c$ equals -38 °C and $T_m$ of -8 °C. This was followed by Zn**NH₄**Ac20 ($T_c$ of -43 °C and $T_m$ -11 °C) and Zn**K**Ac20 ($T_c$ of -50 °C and $T_c$ of -11 °C). The DSC results are consistent with the visual observations after storing the electrolytes at −18 °C for 24 h (Figs. 2b and S9). ZnAc20, Zn**K**Ac20, Zn**NH₄**Ac20, and Zn**Gua**Ac20 remained fully liquid under these conditions, whereas ZnAc underwent complete freezing. However, it is known that the stronger the chaotropic salt is, the lower the transition temperatures of water and aqueous solutions become due to weaker water-water H-bonds.[58] Therefore, the results suggest that the impact of 5MP in the solutions is greater than that of the **Chao⁺** ions, or there may be a synergistic effect between 5MP and the corresponding **Chao⁺** cation. Additionally, the viscosity of the electrolytes (Fig. 2c) slightly increased from 2.8 mPa·s in ZnAc to 3.5-4.0 mPa·s in bi-salt electrolytes, suggesting that the addition of 5MP and bi-salts did not significantly compromise the fluidity of the electrolytes. However, the ionic conductivity (Fig. 2d) was significantly decreased from 17.8 mS/cm in ZnAc to 9 mS/cm in ZnAc20. In contrast, the bi-salts electrolytes exhibited high ionic conductivities and showed an inverse trend to the chaotropic strength of the added ions. Zn**K**Ac20, containing the weakest chaotropic ion ($K^+$), achieved the highest conductivity (28.5 mS/cm), followed by Zn**NH₄**Ac20 (24.2 mS/cm) and Zn**Gua**Ac20 (18 mS/cm). This indicated that 5MP limited the ionic conductivity of the electrolyte, but the bi-salt systems mitigated this effect.[59]

Overall, the results indicate that combining 5MP with chaotropic ions enables the design of a low-concentration electrolyte in which the water environment is significantly disturbed, the operating temperature window is widened, and ion mobility is enhanced, resulting in WiSE-like properties without compromising the physicochemical properties.

**Zn stability**

Cyclic voltammetry (CV) was performed in Cu‖Zn cells at 1 mV/s (Fig. S10) to assess the Zn plating/stripping behavior. Among all electrolytes, ZnAc exhibited the highest initial current density, likely due to multiple parallel surface reactions, but failed by the second cycle, indicating unstable Zn plating/stripping. Zn**Gua**Ac20 showed noticeable shifts after the 50$^{th}$ cycle, while Zn**K**Ac20 displayed minor changes in the 100$^{th}$ cycle. Both observations suggested reduced reversibility and possible side reactions. In contrast, ZnAc20 and Zn**NH₄**Ac20 showed the most stable and reversible Zn plating/stripping behaviour. Linear sweep voltammetry was performed using a stainless steel (SLS) working electrode with an applied low scan rate of 0.2 mV/s to assess slow electrochemical decomposition(Fig. 3a). ZnAc showed a reduction potential of -0.05 V *vs* $Zn/Zn^{2+}$, while all other electrolytes exhibited a reduction potential of -0.08 V *vs* $Zn/Zn^{2+}$. Additionally, the oxidation potential appeared to increase from 2.1 *vs* $Zn/Zn^{2+}$ V in Zn**K**Ac20 to 2.17 V *vs* $Zn/Zn^{2+}$ in Zn**NH₄**Ac and further to 2.2 V *vs* $Zn/Zn^{2+}$ in ZnAc, ZnAc20, and Zn**Gua**Ac20. These results indicate that the addition of 5MP and chaotropic ions had a negligible but positive effect on the electrochemical stability window. The coulombic efficiency (CE) was further determined through a modified Aurbach method (Fig. 3b and S11). While the addition of 5MP increased the CE of ZnAc from 87% to 95% in ZnAc20, the addition of a chaotropic ion did not necessarily translate to higher CE values. Among the chaotropic-containing electrolytes, the Zn**K**Ac20 had the lowest CE value of 89%. Zn**NH₄**Ac20 and Zn**Gua**Ac20 achieved higher efficiencies of 96% and 94%, respectively.

The stability of the Zn anode was further tested at different current densities. First, a C-rate of 1 was used with a capacity of 1 mAh/cm$^2$ (Fig. 3c-d). The symmetric Zn‖Zn cell with ZnAc failed after 135 hours, whereas the addition of 5MP extended the lifetime to 233 hours. Using Zn**K**Ac20 as the electrolyte further extended the cell lifetime to 630 h, nearly a fivefold improvement. A substantially longer lifetime was achieved with Zn**Gua**Ac20 (1670 h), while the longest stability was observed for Zn**NH₄**Ac20, which sustained operation for up to 2000 h. A similar trend was observed under more demanding conditions at 5C (Fig. S12). ZnAc sustained cycling for 168 h, while the addition of 5MP extended the lifetime to 251 h. In comparison, Zn**K**Ac20, Zn**Gua**Ac20, and Zn**NH₄**Ac20 exhibited markedly prolonged lifetimes of 735, 1235, and over 1400 h, respectively. Under these conditions, Zn**NH₄**Ac20 displayed the highest cycling stability among all electrolytes. To isolate the effect of 5MP, Zn**NH₄**Ac without 5MP was evaluated under identical conditions, yielding lifetimes of 185 and 1155 h at current densities of 1 mA cm$^{-2}$ and 5 mA cm$^{-2}$, respectively (Fig. S13). These results highlight the importance of 5MP, particularly at low current densities, in enhancing Zn stability.

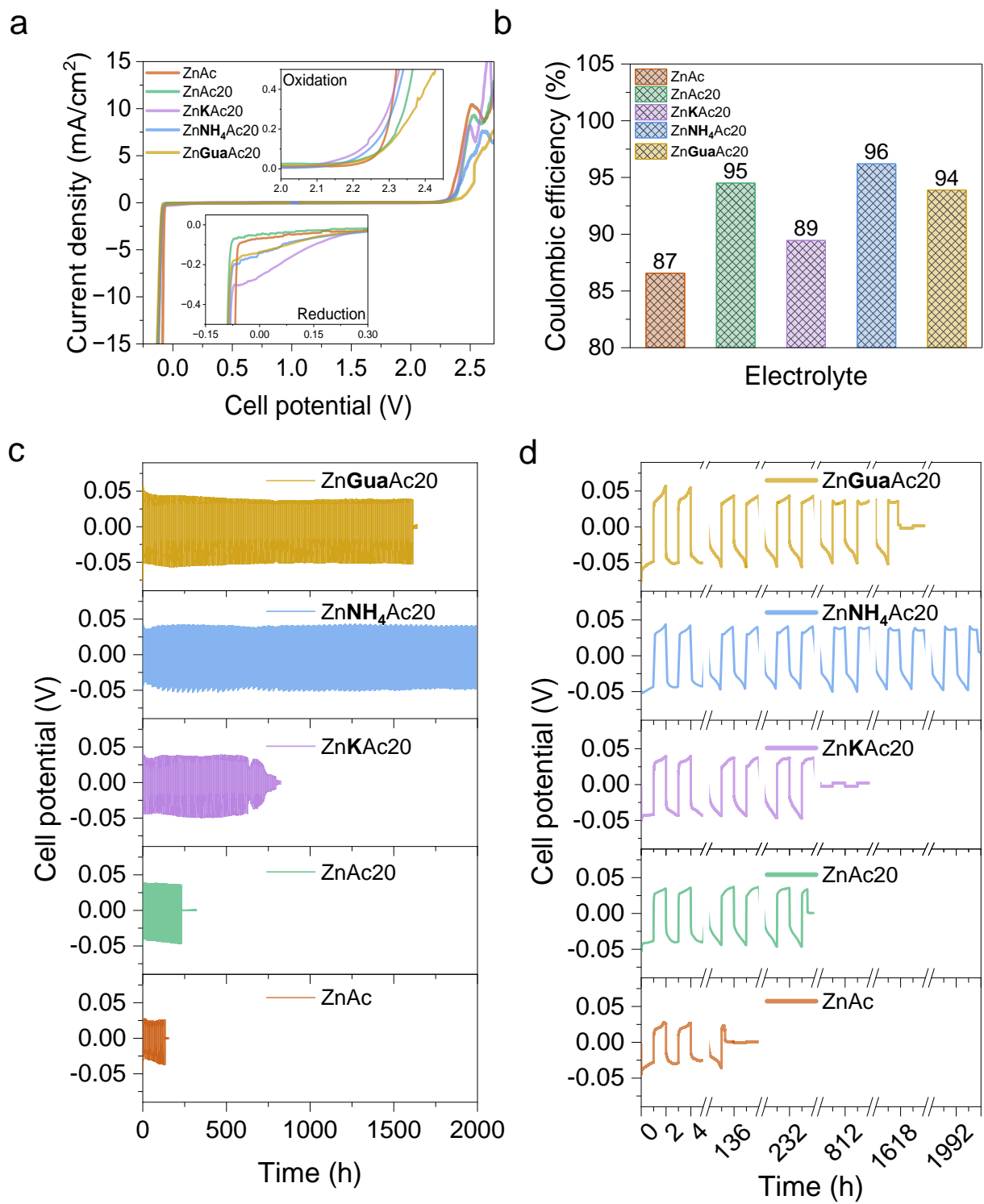

*Fig. 3. Electrochemical stability. a) LSV curves with an applied potential of 0.2 mV/s in an SLS‖Zn cell. b) Determined coulombic efficiency in a Cu‖Zn cell. c-d) Long-term Zn‖Zn cycling at 1 mA/cm$^2$ and 1 mAh/cm$^2$.*

The enhanced cycling stability of the Zn anode observed upon the addition of 5MP to the ZnAc-based electrolyte is primarily attributed to its water-binding capability and its partial replacement of water in the $Zn^{2+}$ solvation sheath, which limits water access to the Zn anode surface and suppresses parasitic side reactions. This is consistent with what Li et al. observed when NMP was used as a cosolvent.[28] Further improvements in anode stability were linked to the chaotropic strength of the added co-salts. It is hypothesised that the weak chaotropic effect of $K^+$ ions was insufficient to bind enough water molecules, resulting in a relatively high free water content compared to the other chaotropic ions.[60] In contrast, the very strong chaotropic effect of $Gua^+$ ions caused a significant disruption of the H-bond in the water network,[61] minimising the amount of free water and possibly increasing the amount of 5MP in the $Zn^{2+}$ solvation sheath. This not only further decreased the water-induced degradation due to fewer water molecules reaching the Zn anode surface, but also restricted $Zn^{2+}$ mobility due to the high viscosity of 5MP, causing a lower ionic conductivity. In contrast, the moderate chaotropic strength of $NH_4^+$ achieved an optimal balance, i.e., enough water molecules were bound to suppress side reactions, while a controlled amount of free water remained to support the $Zn^{2+}$ transport. This balance led to improved ion conductivity and the highest Zn anode compatibility among the tested electrolytes. Therefore, the most effective electrolyte configuration involved the combination of ZnAc, the intermediately strong chaotropic salt $NH_4Ac$, and 5MP. Additionally, these findings highlight that when designing an electrolyte, the aim should not solely be to minimise free water, but rather to fine-tune the balance between bound and free water to ensure both electrochemical stability and efficient ion transport.

To further understand the benefit of 5MP with chaotropic ions, scanning electron microscopy (SEM) was performed after 10 charge/discharge cycles at both 1C and 5C rates, using a fixed areal capacity of 1 mAh/cm². At 1C, all electrolytes showed similar plating morphologies (Fig. S14), but at higher magnifications (Fig. 4a), the differences became more apparent. ZnAc, Zn**K**Ac20, and Zn**Gua**Ac20 showed Zn plated in a more perpendicular orientation on the Zn anode, causing an uneven surface. In contrast, ZnAc20 and Zn**NH₄**Ac20 exhibited more uniformly deposited Zn, resulting in a smoother Zn surface. However, at 5C, only Zn**NH₄**Ac20 showed a relatively flat Zn morphology with few signs of glass fibre (Fig. S15). ZnAc exhibited flat Zn deposition and numerous glass fibres, indicating fibre embedding, likely caused by Zn plating penetrating the glass fibres. ZnAc20, Zn**K**Ac20, and Zn**Gua**Ac20 had similar plated Zn, resembling thin diagonally oriented stripe-like Zn deposits. Additionally, Zn**K**Ac20 also showed cubic-like deposition (as seen in the inset of the SEM image). The X-ray diffraction (XRD) spectra (Fig. 4b-c) further complemented the SEM images, highlighting the preferred Zn deposition orientation of each electrolyte via the relative peak intensities of the (002), (100), and (101) planes. With all bi-salts decreasing the amount of the favourable (002) deposition[62], except Zn**NH₄**Ac20 showed a significant increase in the (002) plane. Based on this performance, Zn**NH₄**Ac20 was selected as the optimal bi-salt electrolyte to further investigate the 5MP and chaotropic effect. Therefore, *in situ* Raman spectroscopy utilising a symmetric Zn∥Zn cell was conducted to investigate further the Zn anode surface in ZnAc and Zn**NH₄**Ac20 electrolytes (Fig. 4d-e and S16). Under a constant current of 1 mA/cm$^2$, the ZnAc exhibited a rapid decrease in the Raman peaks in the water region. Complete surface dehydration was achieved within the first 20 minutes of the stripping phase, indicating water decomposition at the interface.[62] In contrast, Zn**NH₄**Ac20 maintained a consistent water signal intensity throughout the whole measurement, indicating a suppressed hydrogen evolution reaction and improved reversibility of the Zn plating/stripping process.

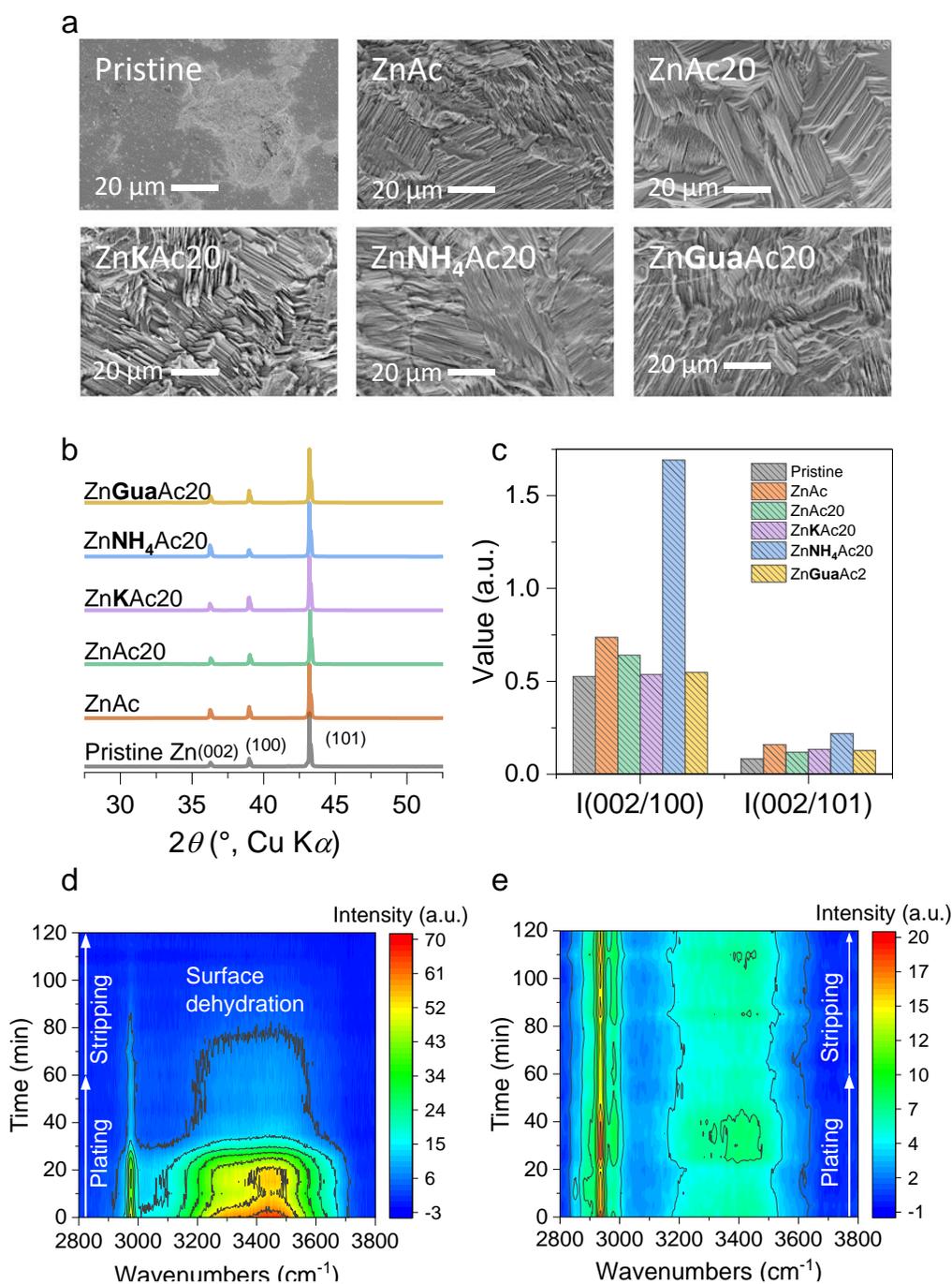

*Fig. 4. Zn anode analysis. a) SEM images after 10 cycles with 1 mA/cm$^2$ and 1 mAh/cm$^2$. b-c) Acquired XRD spectra with peak comparison of the (002), (100), and (101) planes, respectively. d-e) In-situ Raman spectroscopy of Zn surface in ZnAc and ZnNH$_4$Ac20 electrolytes, respectively, during the first cycle at 1 mA/cm$^2$ and 1 mAh/cm$^2$.*

**Performance of 5MP co-solvent in full cell configuration**

In the following, the selected Zn**NH$_4$**Ac20 electrolyte was further tested in full cell batteries using PT-COF as the cathode and compared with ZnAc and ZnAc20 electrolytes. The PT-COF was synthesised by condensation reaction of 1,3,5-triformylphloroglucinol (TFG) and 2,7-diaminopyrene-4,5,9,10-tetraone (DAPT) in mixtures of mesitylene, 1,4-dioxane, and an aqueous acetic acid solution under solvothermal conditions (Fig. S17). The powder X-ray diffraction (PXRD, Fig. S18) and the Fourier transform infrared spectroscopy (FT-IR, Fig. S19) spectra confirmed the successful synthesis of the PT-COF. Simulated PXRD pattern based on an eclipsed (AA) stacking model closely matched the experimental data, and Pawley refinement yielded an excellent fit with minimal discrepancies ($R_p$ and $R_{wp}$ values of 1.34% and 1.78%, respectively). The FT-IR spectrum of PT-COF exhibits two characteristic IR bands at 1252 and 1644 cm$^{-1}$, corresponding to the stretching vibrations of the β-ketoenamine C-N bonds and the C=O groups of the DAPT moiety, respectively. Notably, the N-H stretching bands observed at 3472, 3430, 3370, and 3342 cm$^{-1}$ in the spectrum of the DAPT precursor disappeared in the PT-COF spectrum, indicating the complete consumption of DAPT during the condensation reaction. SEM images of the synthesised PT-COF powder revealed a granular morphology (Fig. S20), while nitrogen adsorption analysis showed a Brunauer-Emmett-Teller (BET) surface area of 341.8 m²/g with a predominant pore size centered at 1.6 nm (Fig. S21a-b). Thermogravimetric analysis (TGA) confirmed that PT-COF is thermally stable up to 200 °C (Fig. S22).

The PT-COF cathode was prepared by mixing the PT-COF with PVDF binder and a conductive carbon additive. The formulation was optimized to achieve the best performance while maintaining a high active-material loading (2.0-3.0 mg/cm$^2$), which is notably higher than what is typically reported for COF-based cathodes in the literature (0.5-0.8 mg/cm$^2$).[49,50] A PT-COF‖Zn battery cell was used for CV experiments to assess the redox activity of the PT-COF material. The initial cycle at 0.5 mV/s (Fig. S23) of ZnAc and Zn**NH$_4$**Ac20 was similar, while ZnAc20 had a low measured current density, indicating less electrochemical activity. However, after 10 cycles, all electrolytes showed the same behaviour where both the redox peaks shifted and moved closer together, most prominently indicating stabilisation of the PT-COF cathode material. Additionally, the cathodic and anodic peaks in the ZnAc electrolyte increased in current density, possibly suggesting more ion insertion, most likely due to less competition between ions to intercalate than in the 5MP and bi-salt electrolytes.[63] The rate capability was tested between 0.1 and 2 A/g, as illustrated in Fig. 5b. At high current densities, all three electrolytes exhibited similar capacities and CE, whereas differences were more pronounced at lower current densities. ZnAc and Zn**NH$_4$**Ac20 were similar in capacities and CE, with both having a CE above 100% at 0.1 A/g, indicating possible inconsistent electrode structure or interactions between the electrolyte and PT-COF at low current densities, which may have caused an inaccurate CE determination.[64] However, Zn**NH$_4$**Ac20 stabilised quickly at 0.2 A/g, indicating that possible parasitic side reactions only occurred at very low current densities. In contrast, ZnAc exhibited the same inconsistency (>100% CE) up to 2 A/g, indicating poor compatibility of ZnAc alone with PT-COF. Additionally, ZnAc did not recover its capacity after cycle 48, while Zn**NH$_4$**Ac20 continued to show stable cycling when reverting to low current densities. ZnAc20 had a lower CE of 96-98% at 0.1 A/g, indicating poor kinetics. However, it had the highest capacity throughout the whole experiment. It also recovered at 0.2 A/g, the same as Zn**NH$_4$**Ac20, and continued to remain stable, showcasing that 5MP positively influenced the stability of the electrolyte-electrode interactions. Examining the profiles of the rate capability (Fig. S24), a plateau was observed for both the charge and discharge phases, corresponding to the anodic and cathodic peaks in the CV results, respectively.

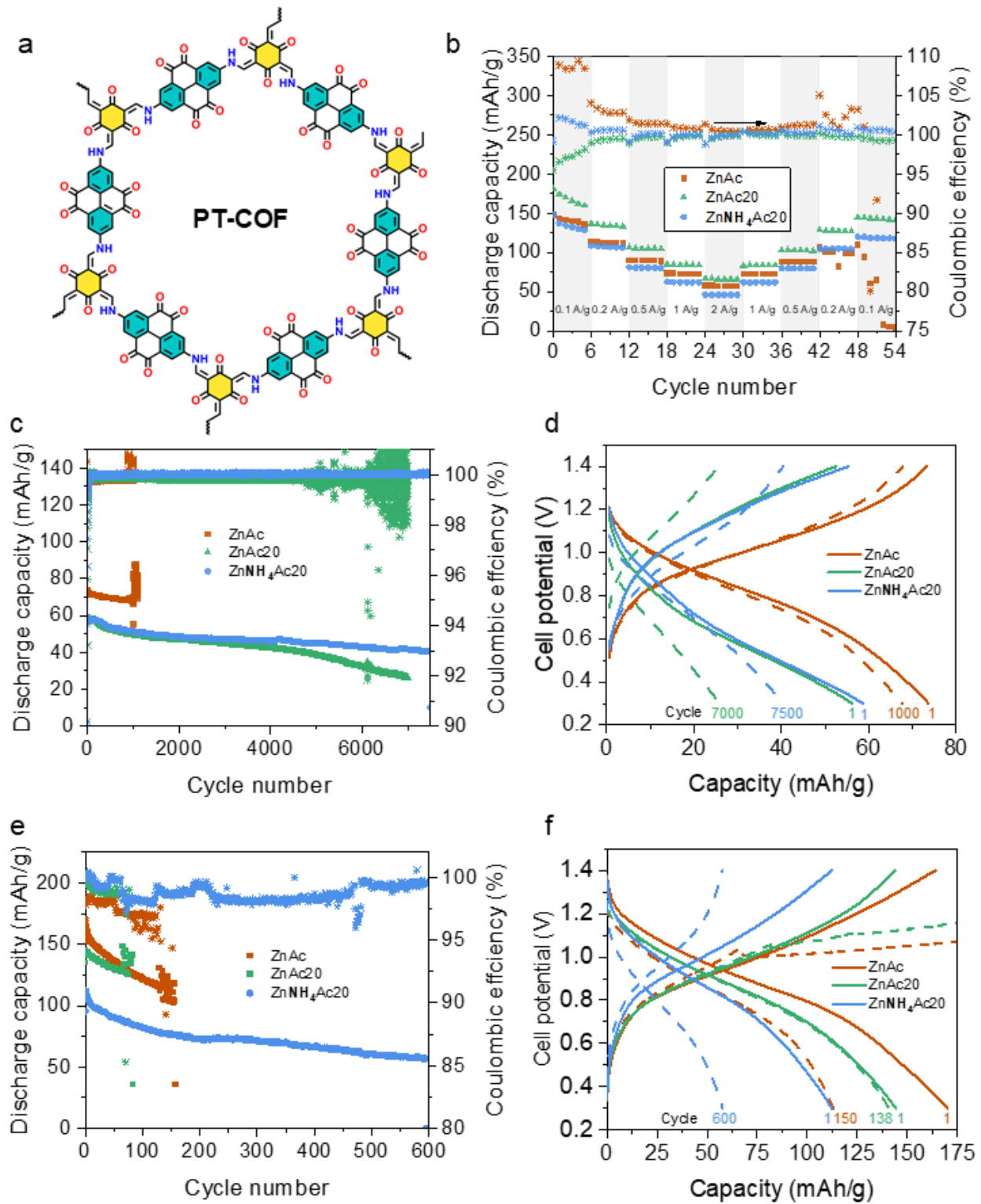

*Fig. 5: COF-based cathode stability. a) Schematic representation of PT-COF. b) Rate capability result. c-d) Long-term stability with an applied current density of 2 A/g. e-f) Long-term stability with an applied current density of 0.1 A/g.*

Afterwards, the current density was kept at 2 A/g (Fig. 5c-d and S25) for the long-term stability test. The ZnAc electrolyte had the highest initial capacity of 73 mAh/g with a high CE of 99.6%, but it could only cycle for roughly 1000 cycles. The ZnAc20 and ZnAc**NH₄**20 exhibited similar results, with Zn**NH₄**Ac20 having slightly higher CE (99.9% and 99.7%, respectively). However, after 4000 cycles, ZnAc20 started to lose capacity more rapidly than ZnAc**NH₄**20, along with unstable cycling as seen from the calculated CE. Zn**NH₄**Ac20 remained stable for over 7200 cycles with a capacity retention of 69%, showing an increase of sevenfold in lifetime compared to ZnAc. As an additional stability test for the PT-COF cathode, 0.2 A/g was utilised to compare the three electrolytes (Fig. S26). All three electrolytes had similar initial capacities, with ZnAc20 having the highest of 130 mAh/g, followed by ZnAc with 123 mAh/g and Zn**NH₄**Ac20 with 116 mAh/g. ZnAc had a total lifetime of fewer than 100 cycles, while ZnAc20 and Zn**NH₄**Ac20 cycled for 335 and 395 cycles, respectively. Even though the Zn**NH₄**Ac20 continued to cycle until 468 cycles, side reactions became apparent from cycle 395 due to the unstable CE. To further assess the stability of PT-COF material at low C-rate, the current density was further decreased to 0.1 A/g (Fig. 5e-f). The ZnAc and Zn20 cycled for roughly 100 cycles with an average CE of 97% and 98%, respectively. The Zn**NH₄**Ac20 cycled for over 600 cycles with a capacity retention of 51% and an average CE of 98%. Despite having the lowest initial capacity (122 mAh/g), Zn**NH₄**Ac20 compensated for it with a significantly longer lifetime. The potential profiles at 0.1 A/g (Fig. S27) were similar in all three electrolytes, suggesting unaltered charge/discharge mechanisms. Additionally, a conventional 2 M Zn(OTf)₂ electrolyte was tested at 0.1 A/g (Fig. S28), but it did not cycle for more than 13 cycles, further showcasing the performance of the developed hybrid electrolyte. Lastly, another cathode material was tested at 2 A/g, namely sodium vanadium oxide (NaVO, Fig. S29). Both ZnAc and Zn**NH₄**Ac20 cycled for over 1150 cycles, but ZnAc had a lower capacity retention of 27% while Zn**NH₄**Ac20 had a retention of 55%. Additionally, ZnAc exhibited a CE above 100% throughout the whole experiment, indicating sluggish kinetics and therefore that not all stored charges were released. Zn**NH₄**Ac20 exhibited a CE of 98-99%, demonstrating its compatibility with NaVO-based cathodes. Although the capacity fade of Zn**NH₄**Ac20 was higher in NaVO-based electrodes than in COF-based ones, the data collectively indicate that Zn**NH₄**Ac20 remains compatible with both cathode materials. This further highlights the advantage of using 5MP as an organic cosolvent in electrolyte formulations to enhance the performance of AZBs.

## Conclusion

A safe and practical organic co-solvent, which can be made from biomass, 5MP, was combined with chaotropic salts (**K**Ac, **NH₄**Ac, **Gua**Ac) to formulate a new class of 1 m ZnAc-based electrolytes, with tunable water structure and improved transport properties. Raman spectroscopy showed a progressive reduction of the overall hydrogen-bond network, with weak H-bonds disrupted preferentially, even at low 5MP contents (10–20 vol%). $^1$H NMR further revealed strengthened N–H···O interactions in 5MP upon salt addition, accompanied by weakened water-water and water-acetate interactions, indicating a strong anchoring role of 5MP while chaotropic ions disrupt the water network. The electrochemical tests demonstrated that a stronger chaotropic effect does not necessarily result in superior Zn-metal reversibility. Among the tested systems, Zn**NH₄**Ac20 provided the best balance of solvation structure and interfacial stability, enabling uniform Zn deposition, wide temperature tolerance, and excellent reversibility. Zn**NH₄**Ac20 delivered over 2000 h of cycling in symmetric Zn||Zn cells and outstanding full-cell stability (>7200 h at 2 A g$^{-1}$, CE > 99%) with PT-COF cathodes. These results demonstrate that a tailored combination of 5MP and NH₄$^+$ chaotropic ions can substantially extend the lifetime and performance of AZBs, offering a robust electrolyte design strategy for next-generation aqueous batteries.

*Electronic Supplementary Information (ESI)*

# Supporting Information for

# Green and Sustainable Hydrogen-Anchored Solvent Enabling Stable Aqueous Zinc Batteries

I. Al Kathemi[a], J. Caroni[b], T. Dehne[a], M. Souto[b,c], M. Antonietti[a], and R. Bouchal[a*]

[a] *Department of Colloid Chemistry, Max-Planck Institute of Colloids and Interfaces, Am Mühlenberg 1, 14476 Potsdam, Germany*

[b] *Centro Singular de Investigación en Química Biológica e Materiais Moleculares Departamento de Química-Física Universidade de Santiago de Compostela, Santiago de Compostela 15782, Spain.*

[c] *Oportunius, Galician Innovation Agency (GAIN), Santiago de Compostela, 15702 Spain*

corresponding author: roza.bouchal@mpikg.mpg.de

## Contents

## Materials

Zinc acetate (ZnAc, 98%) and potassium acetate (KAc, 99%) were obtained from Thermo Scientific. Guanidinium acetate (GuaAc, ≥99%), Zinc trifluoromethanesulfonate (Zn(OTf)$_2$, 98%) were acquired from Sigma Aldrich, but only Zn(OTf)$_2$ was kept and weighed inside the glovebox. Ammonium acetate anhydrous (NH$_4$Ac, 98%) was purchased from MP Biomedicals. N-methyl-2-pyrrolidone (NMP) (99.5%) was supplied by Alfa Aesar for the preparation of the cathode material. Super P carbon black (>99%) was provided by Thermo Scientific, and polyvinylidene fluoride (PVDF) (≥99.5%) and copper foil (99.95%, 10 µm) were purchased from MTI. Zinc foil (99.95%, 100 µm thickness) was purchased from Goodfellow. An alumina polishing suspension (0.05 µm) from Allied High Tech Products Inc. was used to clean the zinc foils, followed by sonicating in acetone for 5 minutes. Carbon paper (200 µm) was provided by Caplinq. 2,7-Diaminopyrene-4,5,9,10-tetraone (DAPT) was obtained from BLD Pharmatech, while 1,3,5-tris(4-formylphenyl)benzene (TFP) was purchased from ET Co., Ltd. Ultra-pure water (18 µS/cm) was used in all experiments.

## 5-methyl-2-pyrrolidone synthesis

5-methyl-2-pyrrolidone (5MP, 99%) was synthesized via the Leuckart reaction by mixing levulinic acid with ammonium formate, following the procedure optimized in our department and described in this reference[1]. 1H NMR (400 MHz, DMSO) δ 7.72 (s, 1H), 3.61 (q, J = 6.1 Hz, 1H), 2.12 (p, J = 4.3 Hz, 3H), 1.48 (tdq, J = 8.0, 5.8, 2.3 Hz, 1H), 1.08 (d, J = 6.2 Hz, 3H).

## Electrolyte preparations

All the electrolytes were obtained by first mixing 5MP in the desired vol% with ultra-pure water. Then, the desired amount of Zn(Ac)$_2$, KAc, NH$_4$Ac, and GuaAc was weighed to achieve 1 m (molality) of each salt. The electrolytes were left stirring at room temperature until the salts were fully dissolved.

## Cathode synthesis and preparation

1,3,5-triformylphloroglucinol (TFG) (10.5 mg, 0.05 mmol), 2,7-diaminopyrene-4,5,9,10-tetraone (DAPT) (21.9 mg, 0.075 mmol), mesitylene (0.2 mL), 1,4-dioxane (0.8 mL), and aqueous acetic acid (200 µL, 6 M) were added to a 10 mL Pyrex tube. This reaction mixture was homogenised by sonication for 15 minutes, and the Pyrex glass tube was subjected to three freeze-pump-thaw cycles and evacuated to an internal pressure of 100 mTorr. The tube was sealed off and then placed in an oven at 120 °C for 3 days. The black precipitate was collected by filtration and washed with DMF, DMSO, and acetone. The resulting solid was dried and then subjected to Soxhlet extraction with methanol as the solvent for 24 hours to remove the trapped guest molecules. The powder was collected and dried under reduced pressure at 85 °C to afford 28.2 mg of PT-COF as a black powder (95% yield).

The obtained COF powder was mixed with Super P carbon black and PVDF in a mass ratio of 6:3:1. NMP was slowly added during stirring until a slurry of desired consistency was acquired. The slurry was left to stir overnight before casting on carbon paper with a wet thickness of 150 µm. The carbon paper with the COF slurry was left to dry overnight in an oven at 60 °C with a vacuum before being manually punched into disk-shaped cathodes (Ø 6 mm, loading 2-3 mg/cm$^2$).

The NaVO material was synthesised in the same method as stated in our previous publication[2] and based on the method of Wan et.al.[3]. In short, NaV$_3$O$_8$·1.5H$_2$O (NaVO) was synthesized by weighing 1 g V$_2$O$_5$ and mixing it with 15 mL 2 M NaCl and stirring for 96 h at 30°C until the solution turned dark red, indicating

Na⁺ insertion and nanorod formation[3,4]. The product was washed (water and ethanol, 5 times each) and dried at 60°C overnight.

# Characterisations

## Differential scanning calorimetry

DSC measurements were conducted using a DSC 300 Caliris Select StandardP (Netzsch, Selb, Germany) with ~10 ± 1 mg of sample in an aluminium pan with a pierced lid. Heating/cooling was performed at 10 K/min, and data were processed with Proteus software (v9.6.2).

## Ionic conductivity

Ionic conductivity of the electrolytes was measured by electrochemical impedance spectroscopy using calibrated Pt/C electrodes, with cell constants ($k_C$) between 0.9–1.1. Measurements were carried out at 25 °C over 100 kHz–50 mHz with a 10 mV amplitude. Ionic resistance was obtained from the intercept of the real axis.

## Proton Nuclear Magnetic Resonance

$^1$H NMR was conducted coaxially on an Agilent 400 MHz spectrometer with DMSO-d6 as a standard reference solvent.

## Raman spectroscopy

Raman spectra were recorded using a WITec Alpha 300M+ in confocal backscattering mode with a 20× objective. A 532 nm laser (50 mW) was used with 0.5 s integration time, 60 accumulations, and 2 cm⁻¹ resolution. Electrolytes were measured in a 2 mm quartz cuvette.

In-situ Raman spectroscopy was performed in an operando cell equipped with a quartz window, using a 50× objective. Each measurement used a 532 nm laser at a power of 50 mW, an integration time of 1 second, and 50 accumulations with a resolution of 2 cm$^{-1}$. Each spectrum is the average spectrum of 5 minutes of continuous measuring.

## Scanning electron microscopy

The morphology of the COF sample was examined by Zeiss Sigma 300 FM SEM. The Zn anode was analysed on a LEO 1550 Gemini Zeiss microscope at 5 kV. Samples were vacuum-dried for 10 min before analysis.

## X-ray diffraction

XRD measurements on the Zn anodes were performed on a Rigaku SmartLab diffractometer using Cu Kα radiation (λ = 1.5406 Å) with a step size of 0.05° and a scan rate of 1°/min.

## Powder X-ray diffraction (PXRD)

PXRD patterns were recorded at room temperature using a Rigaku MiniFlex diffractometer operating in Bragg–Brentano geometry (θ/θ configuration, 40 kV, 40 mA) equipped with a sealed Cu X-ray tube (λ = 1.5406 Å, Cu Kα₁ radiation). Data were collected over the 2θ range of 2–40° with a step size of 0.01° and a counting time of 0.4 s per step. To minimize background contributions, the powder samples were mounted on a Si (511)-oriented crystal substrate instead of conventional glass supports. Pawley refinement was performed on the experimental PXRD pattern after an initial fitting of the diffraction data.

All refinements were carried out using the HighScore Plus software package. The refinement employed a structural model with AA stacking, consistent with previously reported structures for this class of COFs. The background was manually adjusted prior to refinement to ensure accurate peak fitting, and peak profiles were described using a pseudo-Voigt function. Lattice parameters were refined until convergence was achieved, yielding an excellent match between the experimental and calculated patterns (Rp = 1.34%, Rwp = 1.78%).

### Gas sorption measurements (BET)

$N_2$ adsorption-desorption isotherms were measured ex-situ on a Micromeritics 3Flex apparatus at 77 K. Before measurements, the samples were outgassed at 393 K and $10^{-6}$ Torr overnight using Smart VacPrep (Micrometrics) equipment. Pore size distributions were obtained using the non-local density functional theory (NLDFT) method.

### Thermogravimetric Analysis (TGA)

TGA measurement was performed on a TA Instruments Q5000 IR thermobalance. The sample was heated from 25 to 800 °C at a rate of 5 °C min$^{-1}$ under a nitrogen atmosphere (flow rate: 25 mL min$^{-1}$).

### Fourier Transform Infrared Spectroscopy (FTIR)

FTIR spectra were recorded using a PerkinElmer Spectrum Two spectrometer equipped with an attenuated total reflection (ATR) accessory. Dried powder samples were placed directly on the ATR crystal, and spectra were collected over the 400–4000 cm$^{-1}$ wavenumber range.

### Zn anode compatibility

Cycling stability of Zn‖Zn cells was tested in two-compartment Swagelok cells with Ø10 mm Zn electrodes, a Ø13 mm glass fiber separator, and 90 µL electrolyte. After 10 hours of rest, cells were cycled at 1 or 5 mA/cm² with a capacity of 1 mAh/cm². Fresh cells were assembled for SEM comparison. The cutoff potential was 0.5 V.

### Viscosity

Viscosity was measured using an Automated Microviscometer (AMVn, Anton Paar, Graz) based on Hoeppler's falling ball principle. A 1.6 mm internal-diameter capillary was filled with the solution, and a 1.5 mm stainless-steel ball was used. Measurements were performed at a 70° angle, in duplicate, with six repetitions at 25 °C.

## Electrochemical measurements

### Coulombic efficiency

Coulombic efficiency was evaluated using the modified Auerbach method by Vazquez et al.[5] in a two-compartment Swagelok cell with a Ø12 mm Cu foil WE, Ø10 mm Zn foil CE, Ø13 mm glass fibre separator, and 90 µL electrolyte. The protocol consisted of three steps: (i) Zn plating–stripping–replating at 3 mA/cm² with 5 mAh/cm² capacity, (ii) 10 Zn plating/stripping cycles at 0.5 mA/cm² and 1 mAh/cm², and (iii) Zn stripping at 3 mA/cm² with a 0.5 V (vs. Zn/Zn$^{2+}$) cutoff.

### Cyclic voltammetry

The cyclic voltammetry of the Zn anode was determined with a Ø 12 mm Cu foil WE, a Ø 10 mm Zn foil as CE, a Ø 13 mm glass fibre separator and 90 µl of electrolyte. The applied potential was 1 mV/s in a potential window from +0.4 to -0.2 V.

For the COF cathode, the CV was determined in the potential window of 0.4 to 1.6 V, with an applied potential of 0.5 mV/s. The cell consisted of a Ø 6 mm COF cathode as WE, a Ø 8 mm Zn foil as CE, a Ø 13 mm glass fibre separator and 90 µl of electrolyte.

### Cathode compatibility

The rate capability and long-term cycling of the full cells were performed in a COF‖Zn Swagelock cell with a Ø 6 mm COF cathode as WE, a Ø 8 mm Zn foil as CE, a Ø 13 mm glass fibre separator and 90 µl of electrolyte. For the rate capability, the applied current density was varied between 0.1 to 2 A/g. For the GCD cycling, new Swagelock cells with clean electrodes were assembled to test the compatibility with 0.1, 0.2, and 2 A/g.

Figures S1-29

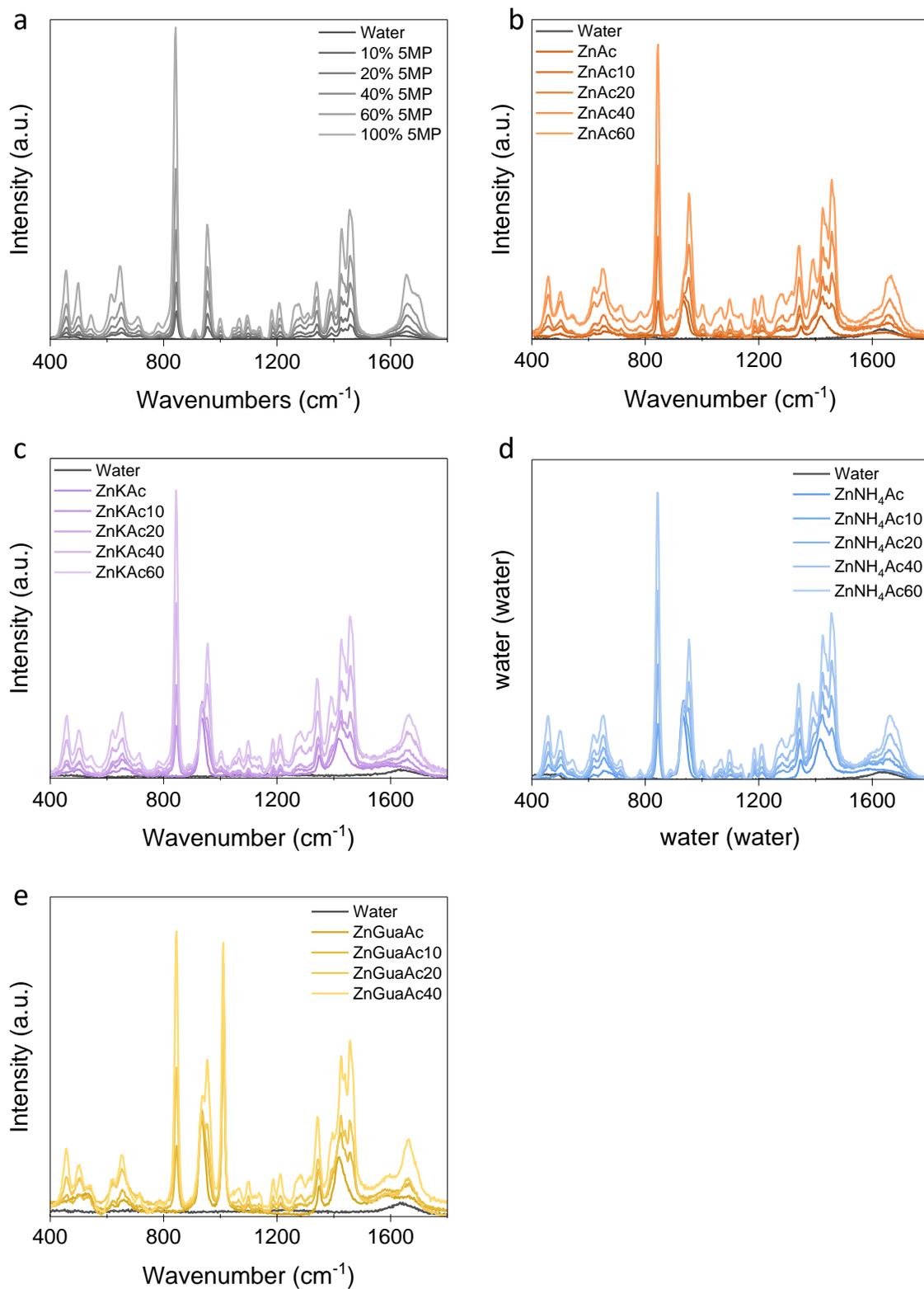

*Figure S1: Raman spectroscopy of the effect of 5MP in a) water, b) ZnAc, c) ZnKAc, d) ZnNH4Ac and e) ZnGuaAc.*

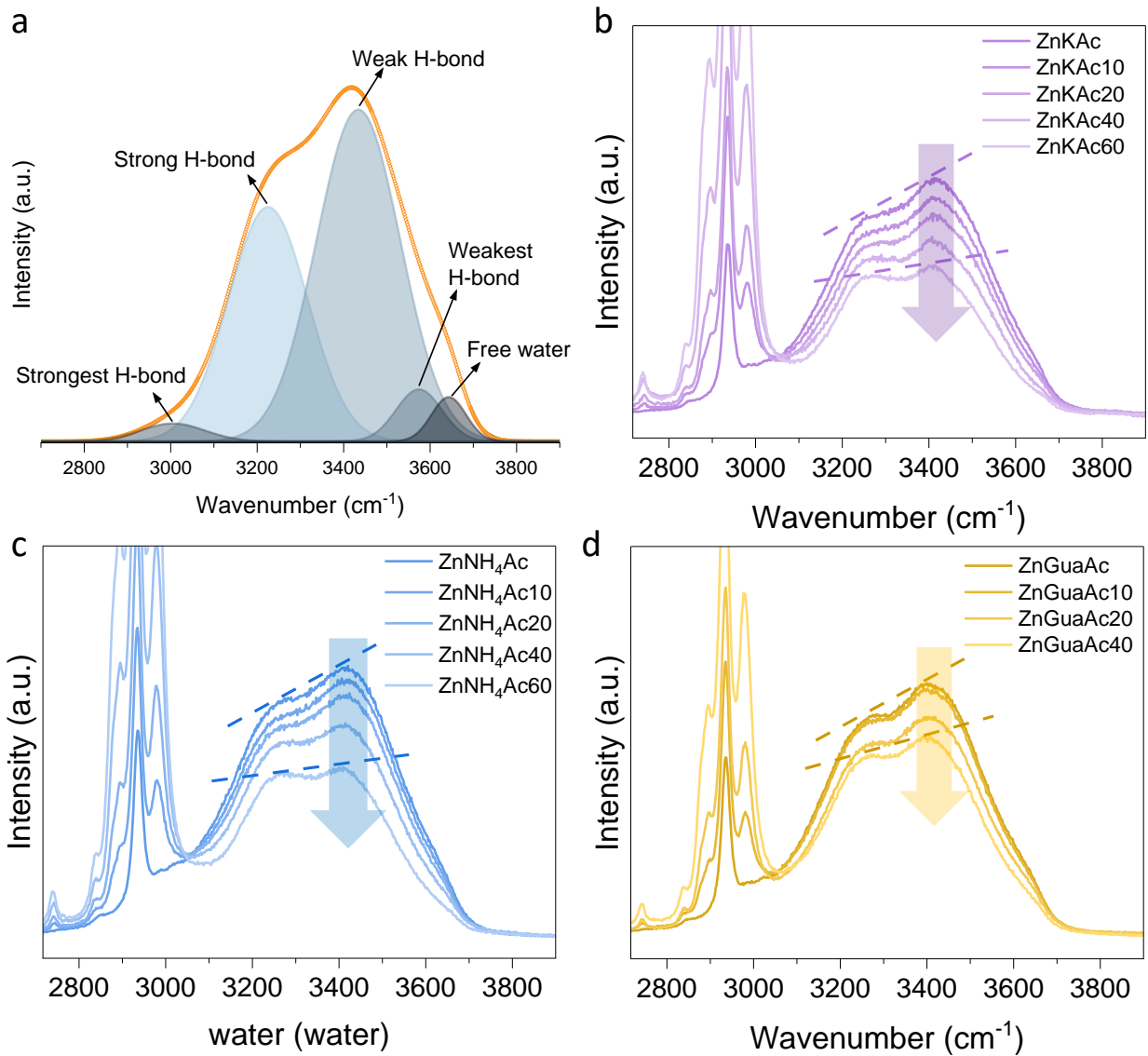

*Figure S2: Raman spectroscopy of the effect of 5MP on the water environment in a) ZnAc, b) ZnKAc, c) ZnNH4Ac and d) ZnGuaAc. e) Deconvolution of the water region of ultra pure water.*

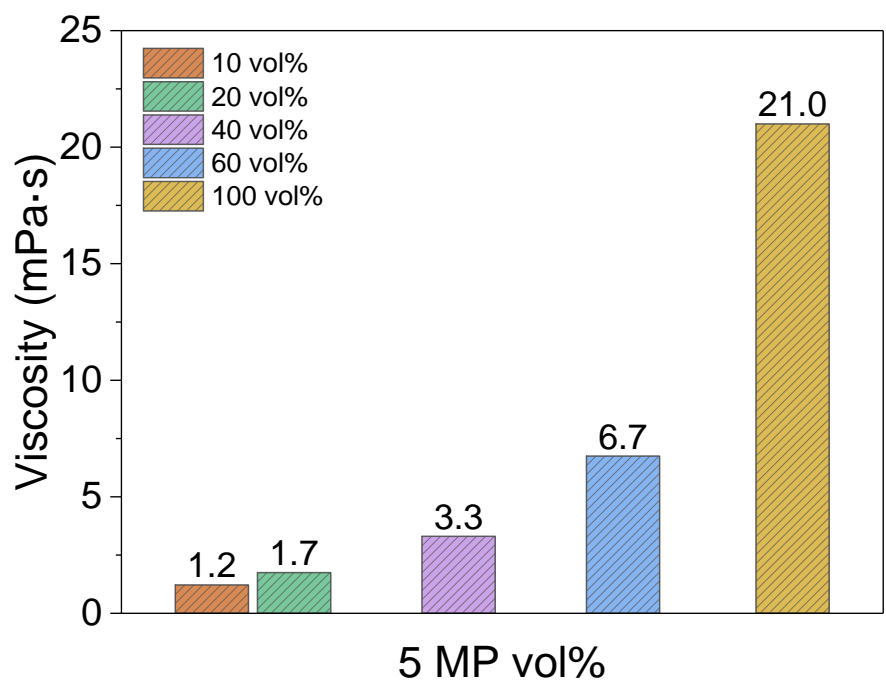

Figure S3: Viscosity of different vol% of 5MP in water.

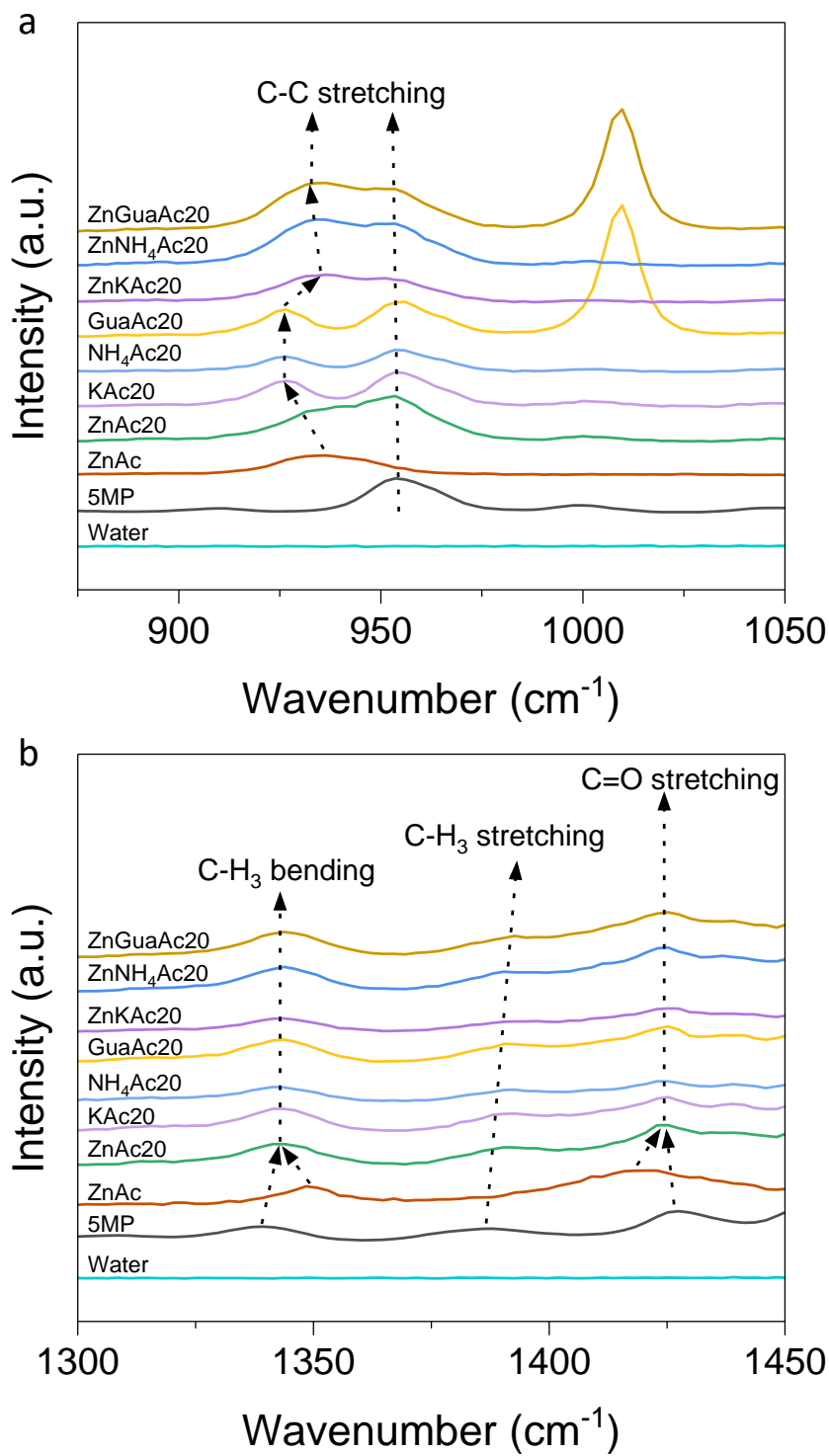

*Figure S4: Raman spectroscopy of the effect of 5MP on different Raman regions. a) From 875 to 1050 cm$^{-1}$. b) From 1300 to 1450 cm$^{-1}$.*

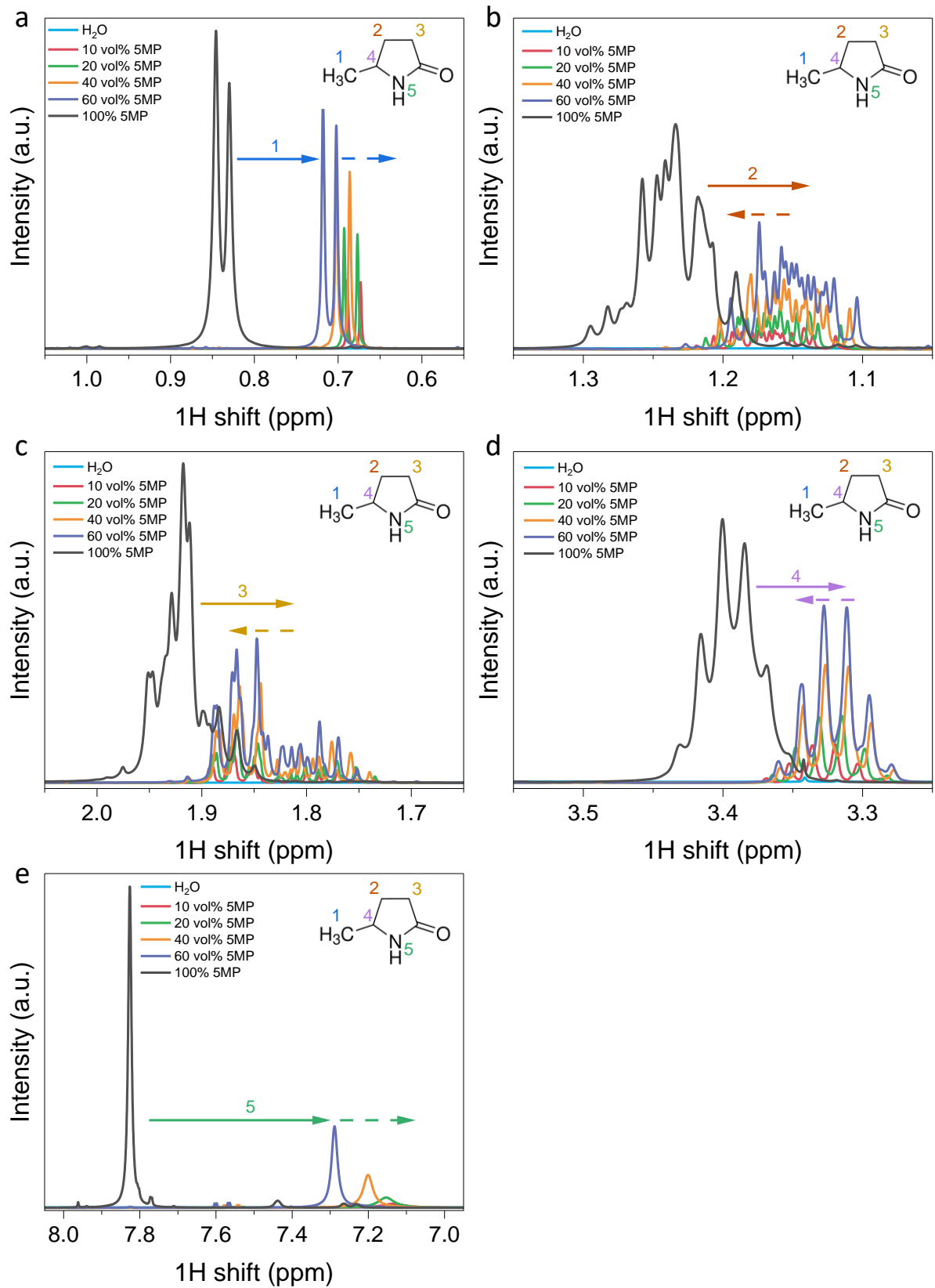

Figure S5: 5MP peak shifts in $^1$H NMR. a) CH3, b) CH2 and c) CH.

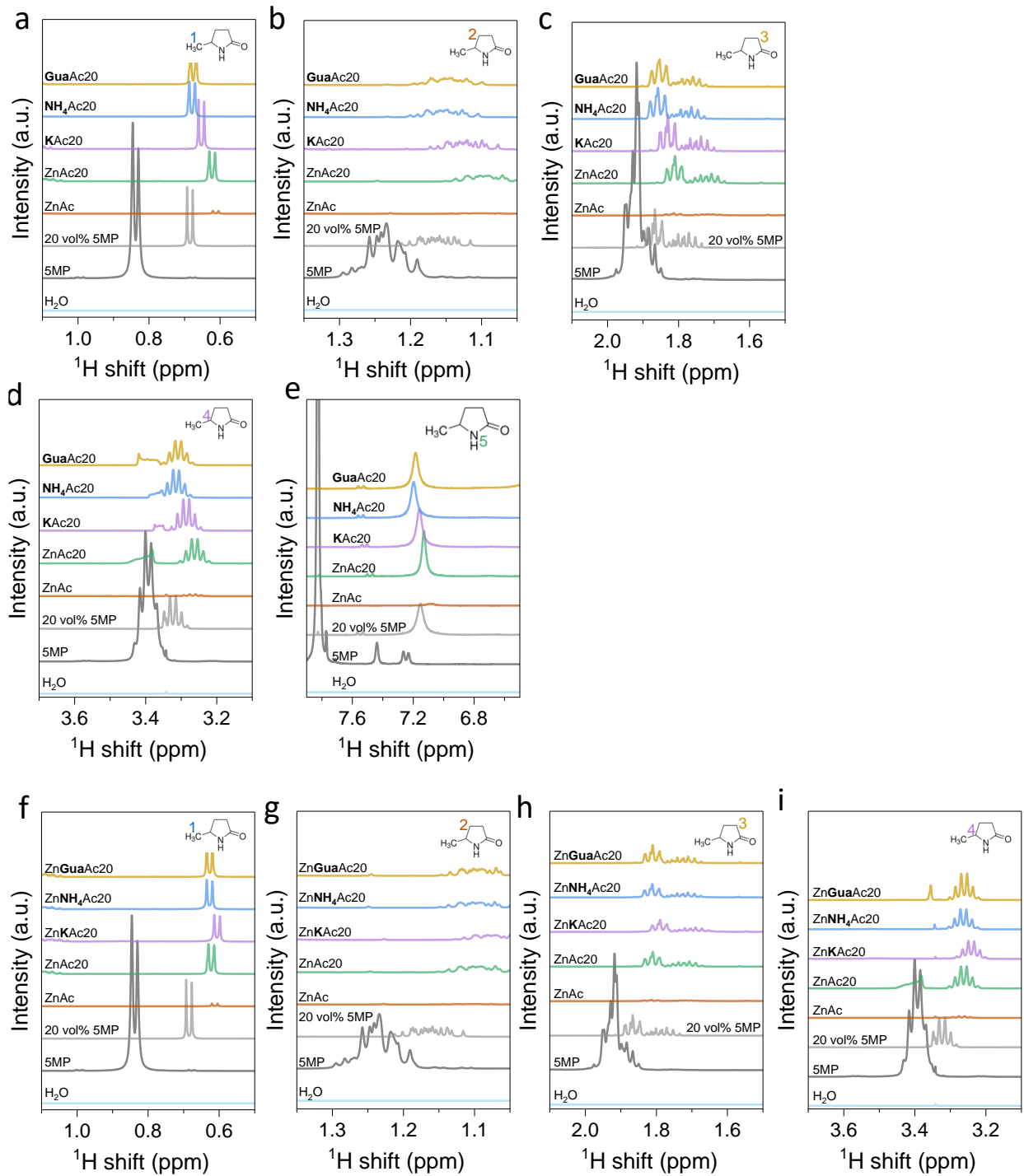

Figure S6: 5MP peak shifts in $^1$H NMR in both single and bi-salts electrolytes with 20 vol% 5MP. a-e) Single salts. f-i) Bi-salts.

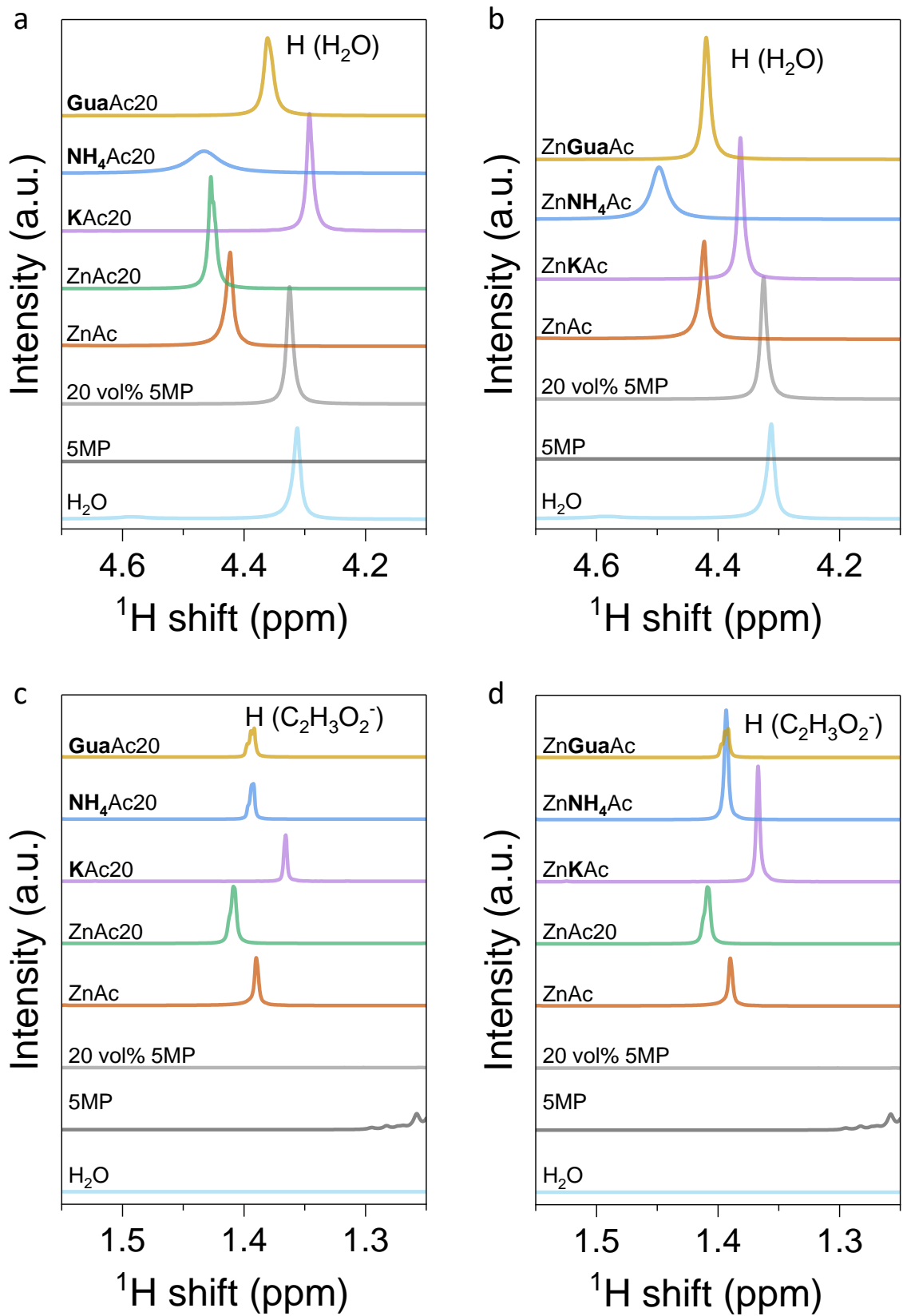

Figure S7: 5MP peak shifts in $^1$H NMR. a-b) Water. c-d) acetate ion.

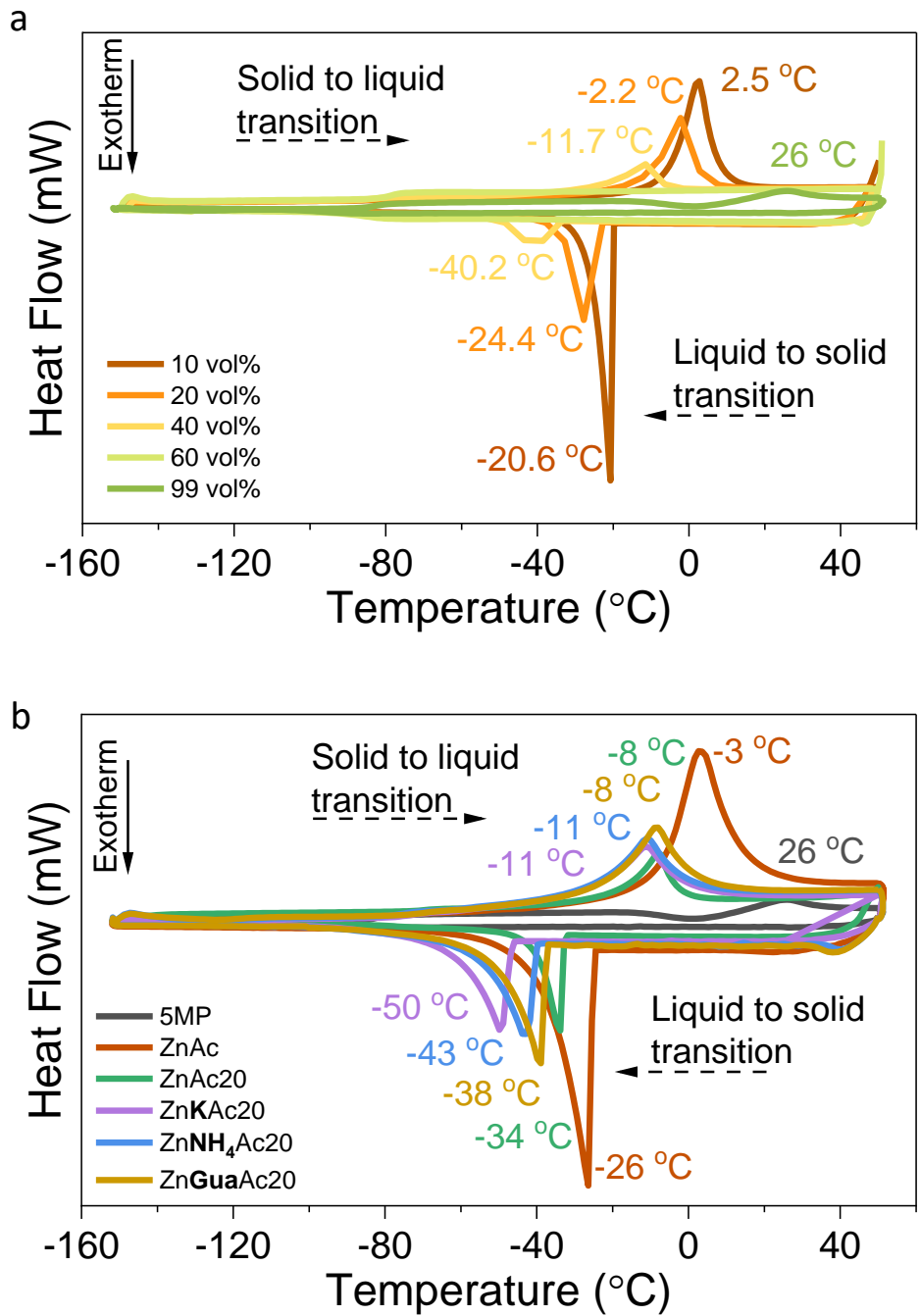

Figure S8: DSC results of a) different vol% 5MP and b) selected electrolytes.

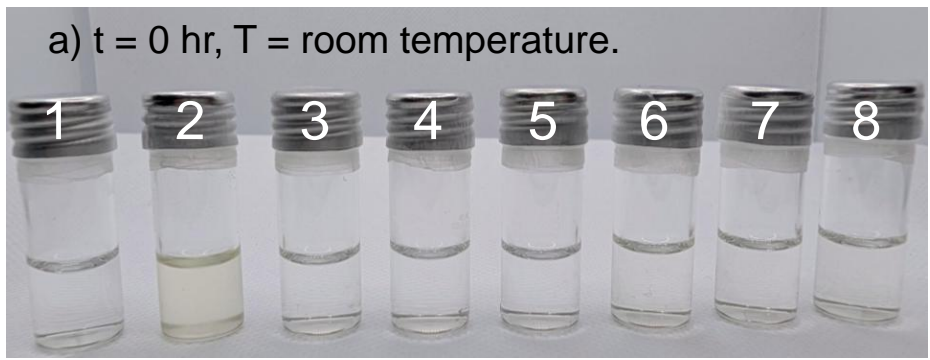

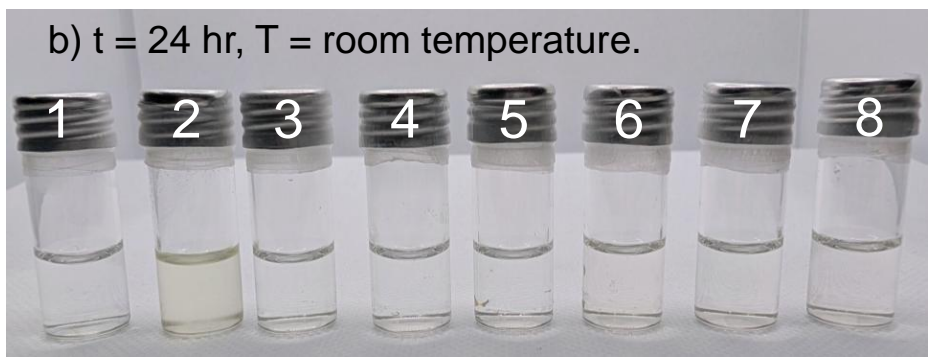

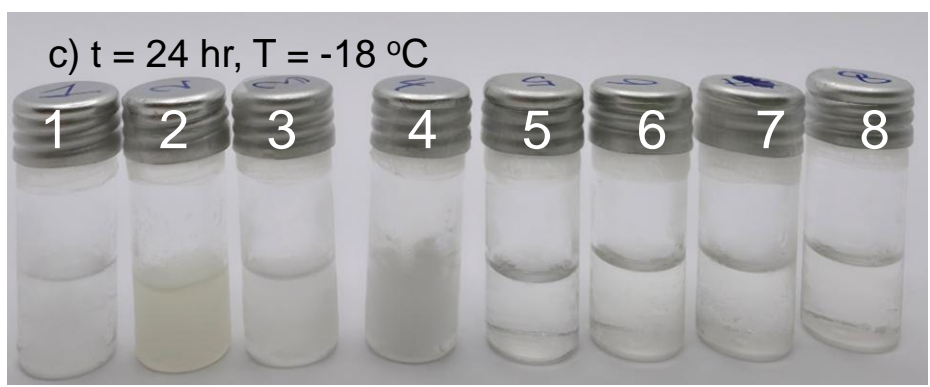

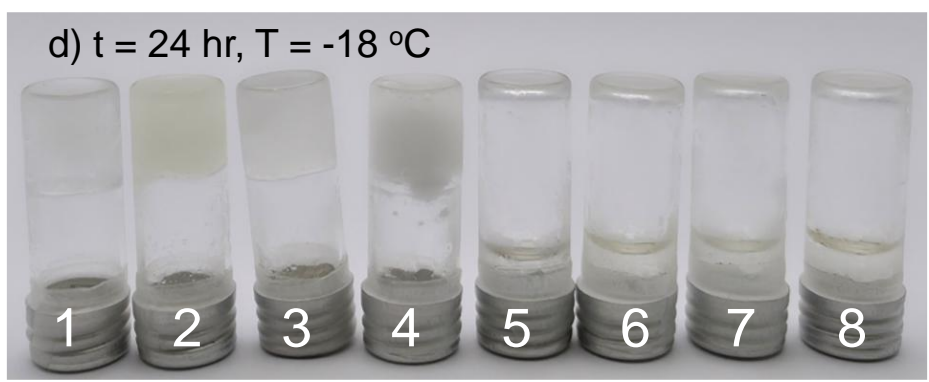

*Figure S9: Digital photograph of the electrolytes. a) Electrolytes after preparation. b) Electrolytes after 24 hours at room temperature. c-d) Electrolytes after 24 hours at -18 °C. The electrolytes are numbered as follows: 1) Ultra pure water, 2) pure 5MP, 3) 20vol% 5MP in ultra pure water, 4) 1 m Zn(Ac)$_2$ in ultra pure water, 5) 1 m Zn(Ac)$_2$ in 20 vol% 5MP solution, 6) 1 m Zn(Ac)$_2$ and 1 m **K**Ac in 20 vol% 5MP solution, 7) 1 m Zn(Ac)$_2$ and 1 m **NH$_4$**Ac in 20 vol% 5MP solution, 8) 1 m Zn(Ac)$_2$ and 1 m **Gua**Ac in 20 vol% 5MP solution.*

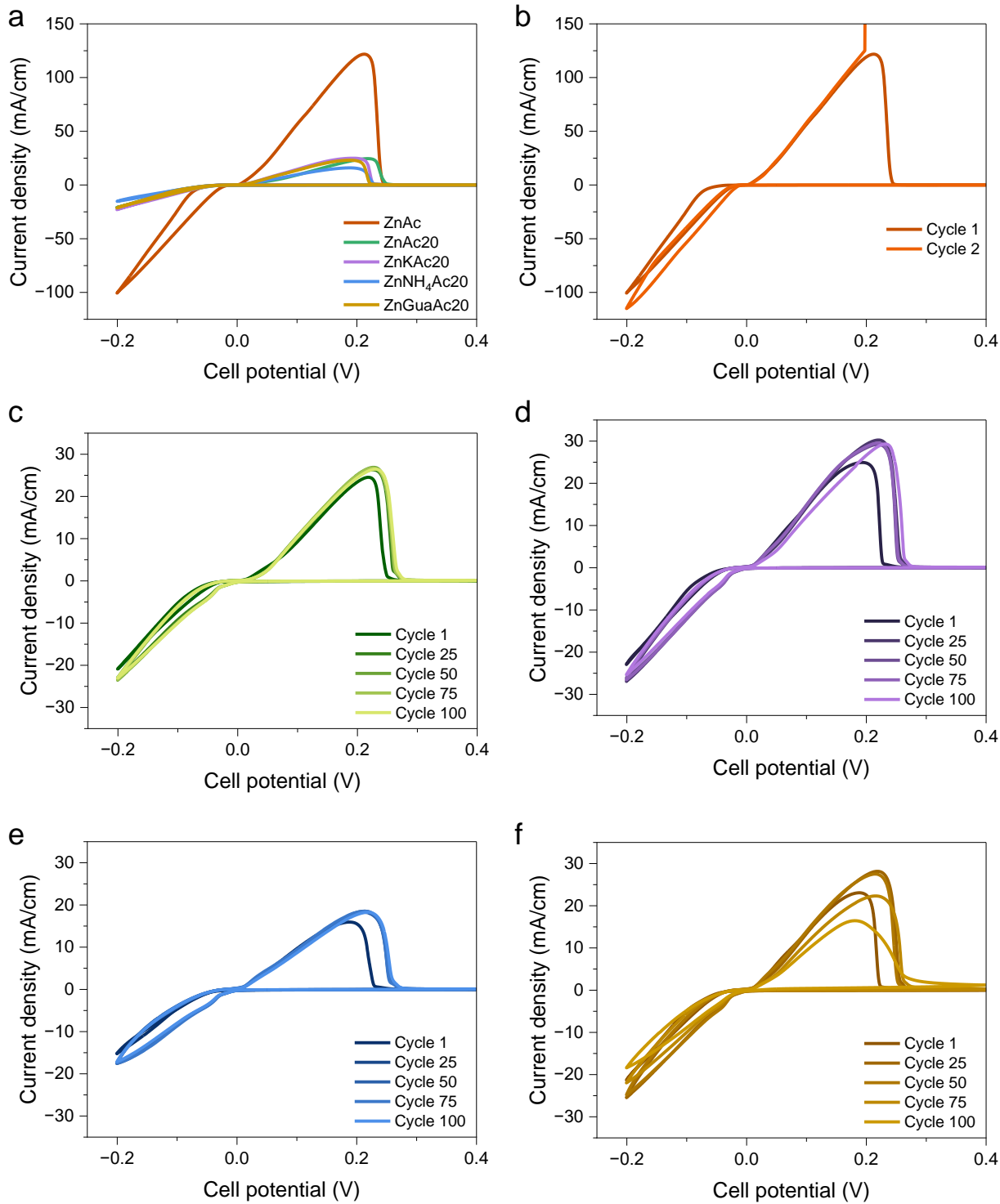

*Figure S10: CV characterisation in a Cu‖Zn cell. a) First cycles of all electrolytes. Long-term CV curves of b) ZnAc, c) ZnAc20, d) Zn**K**Ac20, e) Zn**NH₄**Ac20 and f) Zn**Gua**Ac20.*

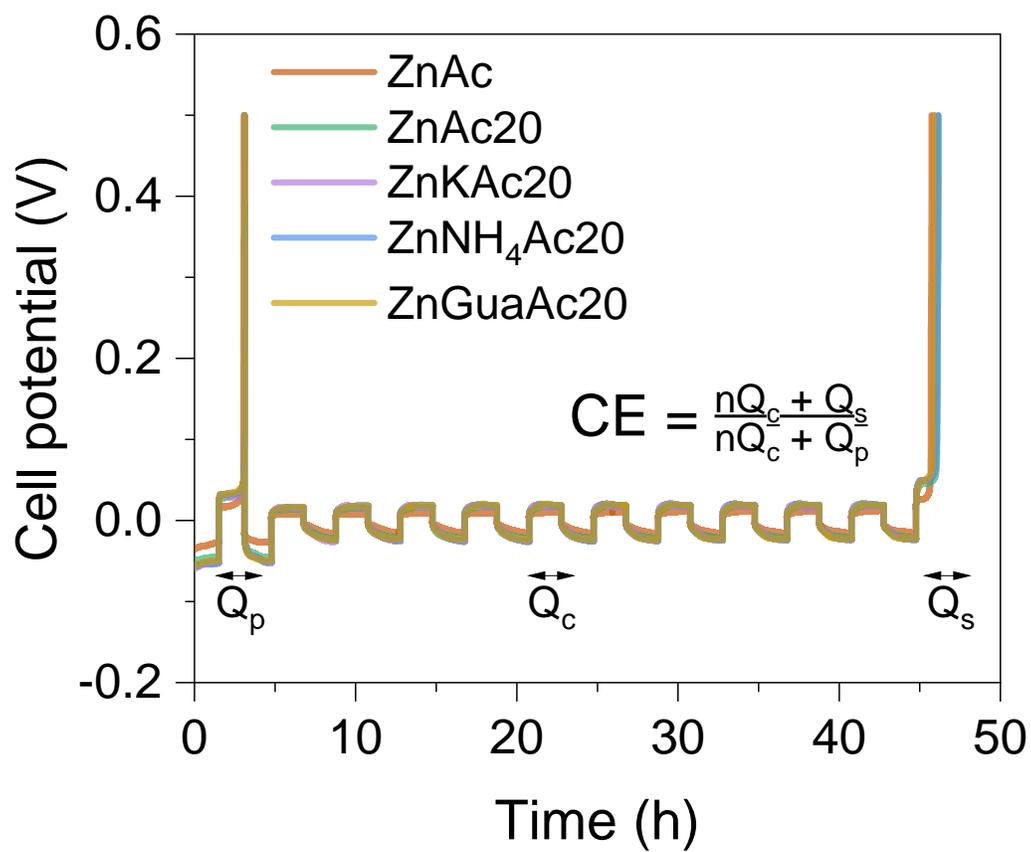

*Figure S11: Potential curves of the modified Aurbach method for determining the CE.*

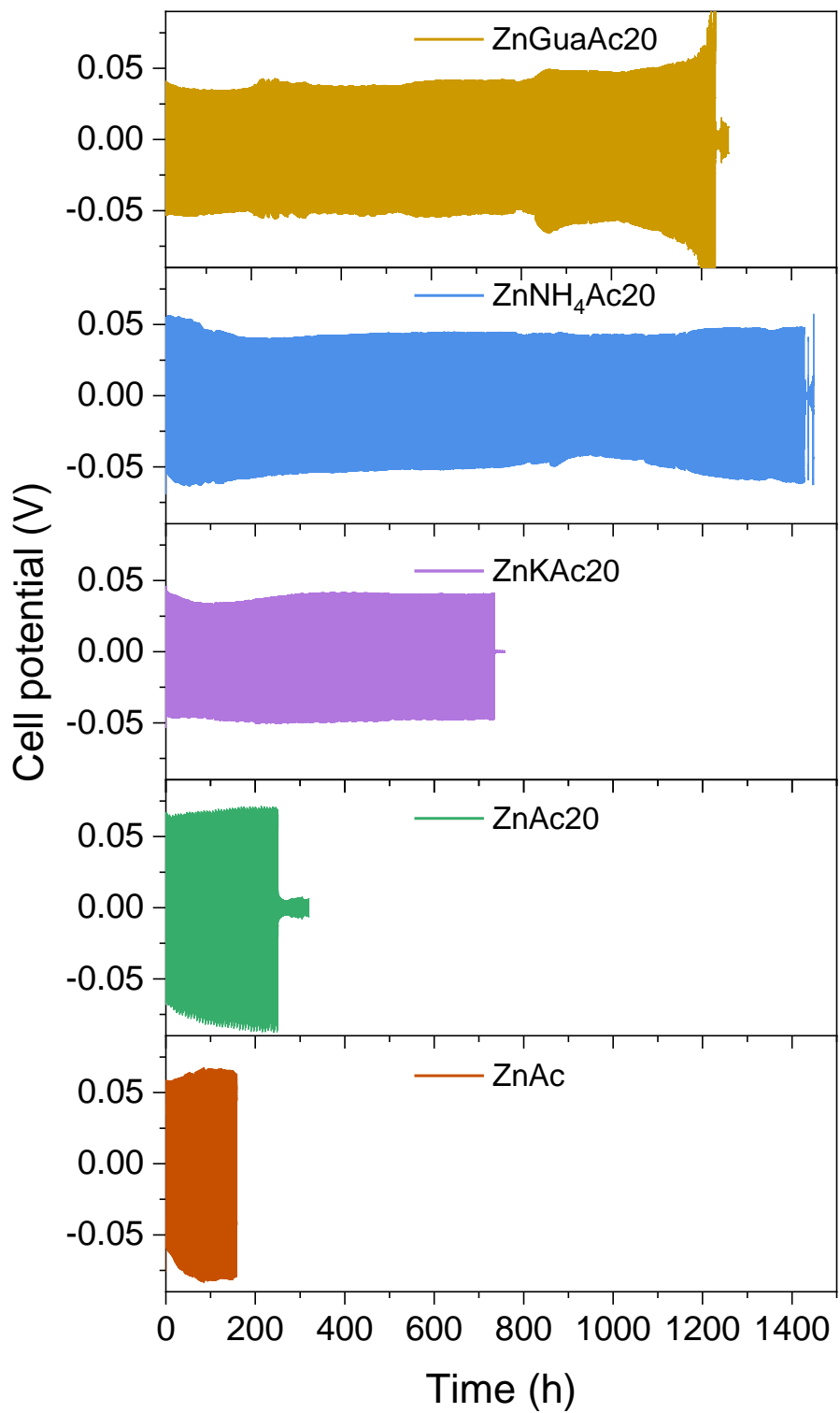

Figure S12: Long-term stability of a Zn‖Zn cell utilising 5 mA/cm² and 1mAh/cm² for the selected electrolytes.

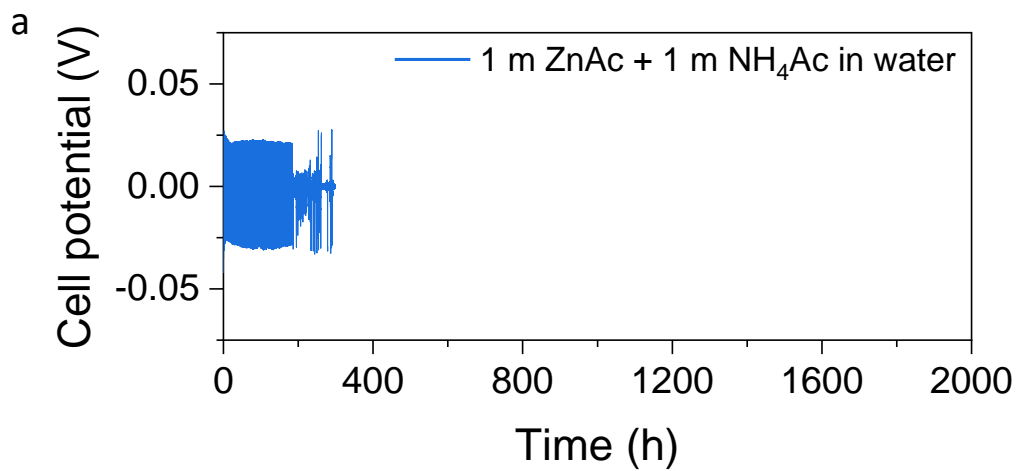

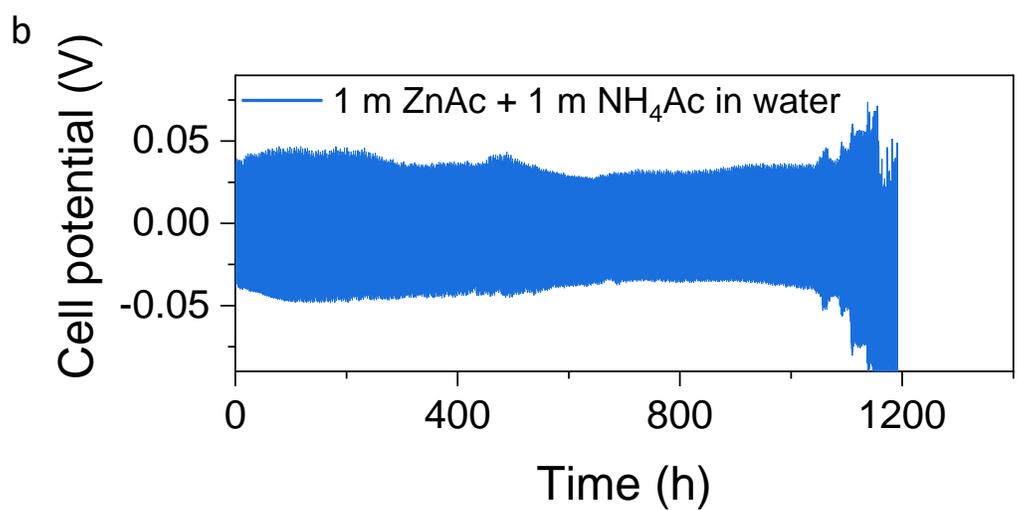

*Figure S13: Long-term stability of a Zn∥Zn cell utilising a) 1 mA/cm² and b) 5 mA/cm² with 1mAh/cm² for the electrolyte 1 m Zn(Ac)₂ with 1 m NH₄Ac in water.*

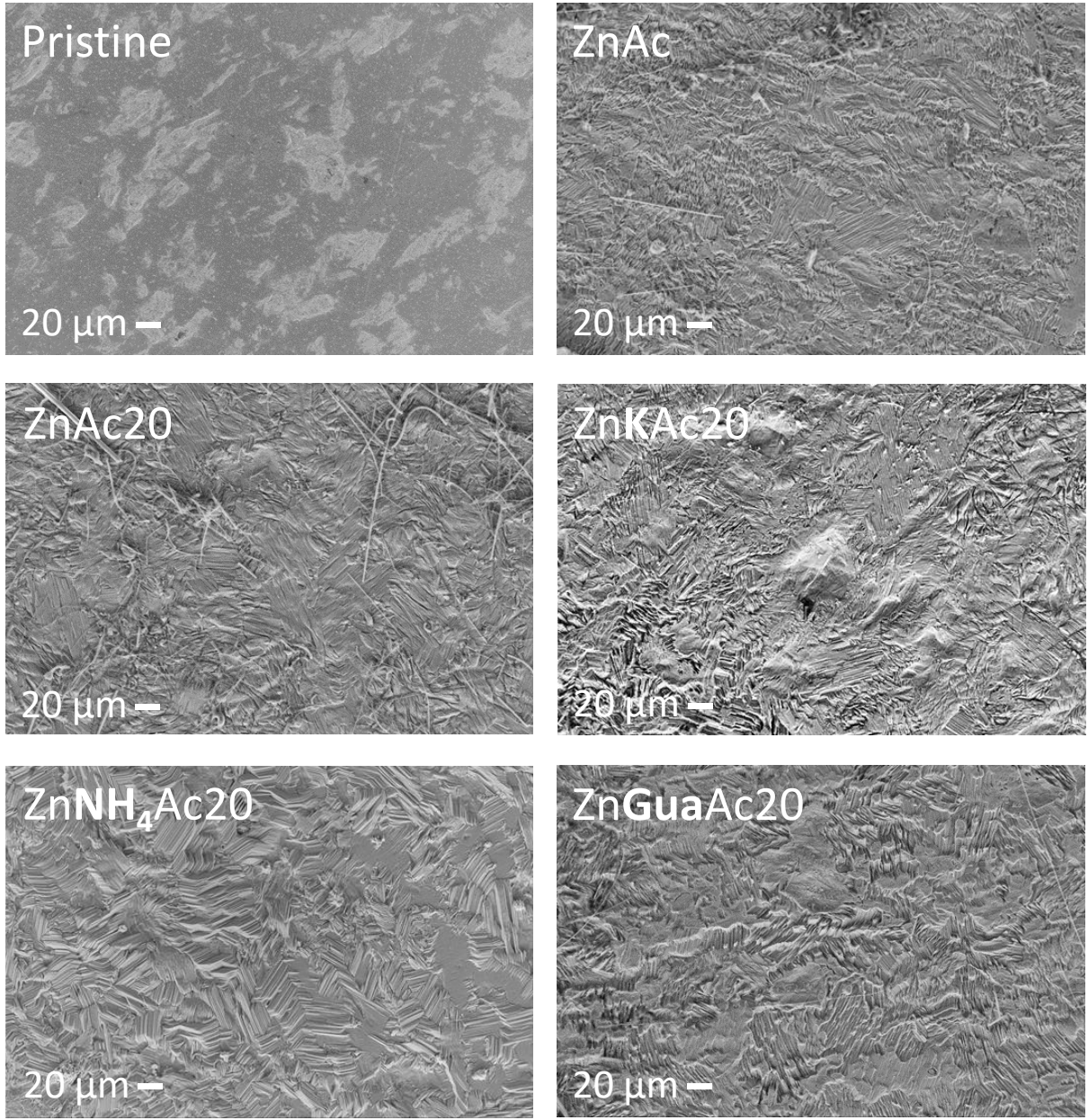

*Figure S14: SEM images of the recovered Zn anodes after applying 1 mA/cm$^2$ with a capacity of 1 mAh/cm$^2$ for 10 cycles.*

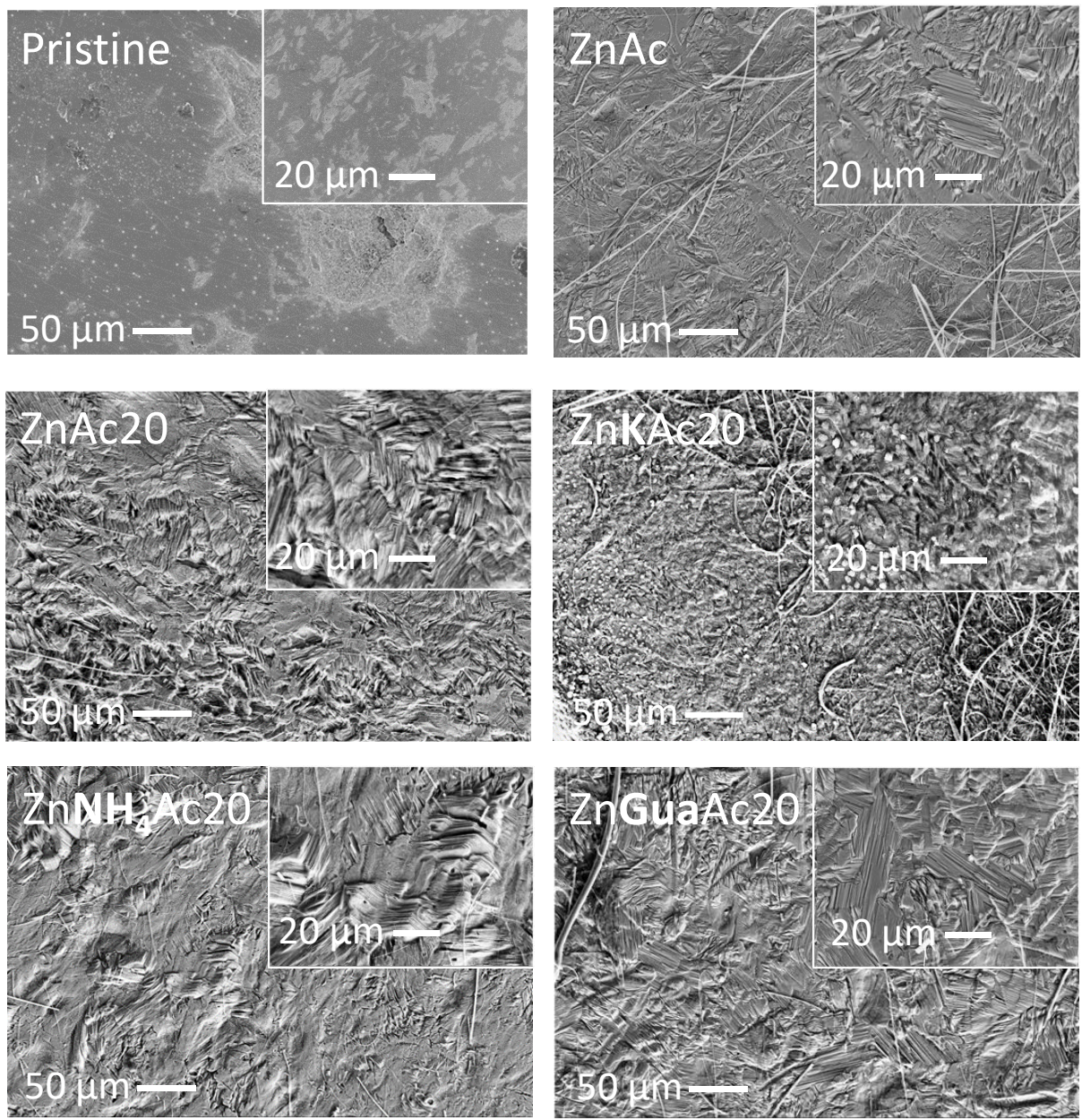

Figure S15: SEM images of the recovered Zn anodes after applying 5 mA/cm² with a capacity of 1 mAh/cm² for 10 cycles.

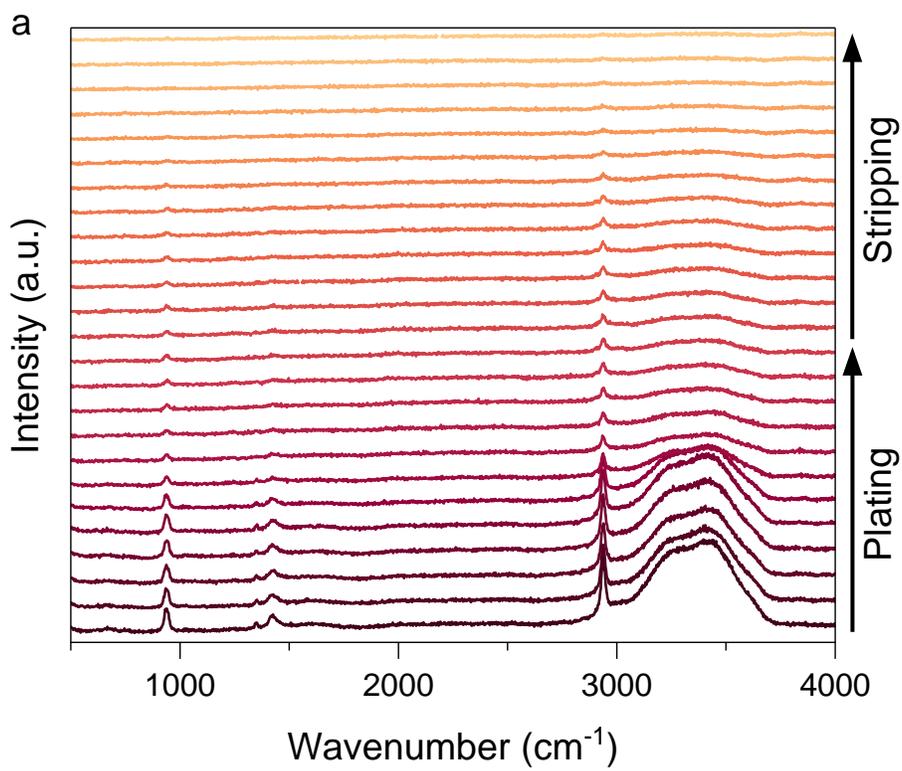
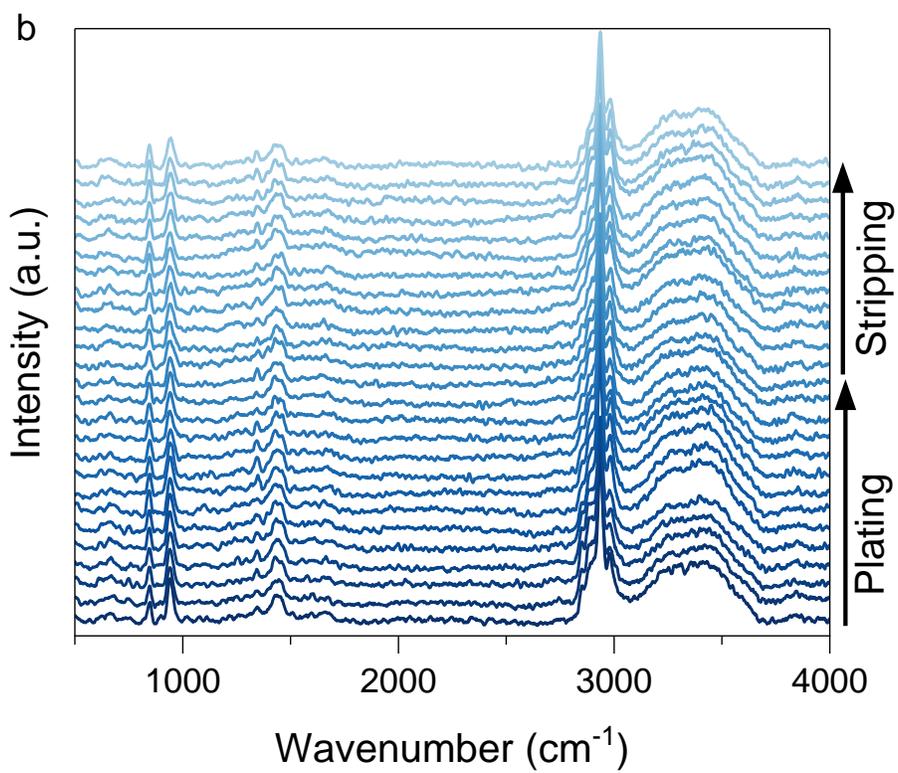

*Figure S16: In-situ Raman spectroscopy during the first charge and discharge cycle in a Zn||Zn cell utilising a) ZnAc and b) Zn**NH₄**Ac20.*

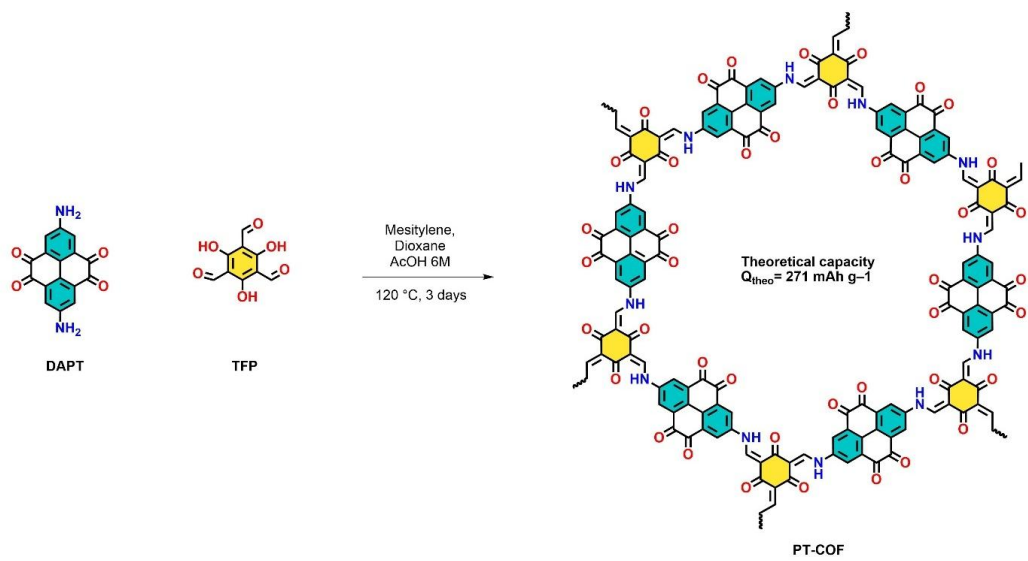

*Figure S17: Synthesis of PT-COF.*

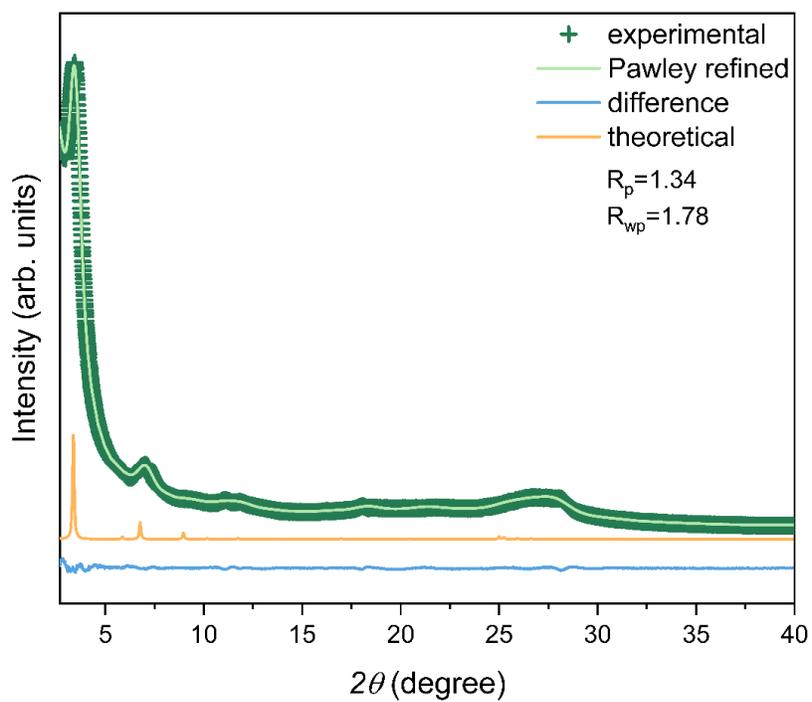

*Figure S18: PXRD pattern of PT-COF with Pawley refinement (Rp = 1.34%, Rwp = 1.78%).*

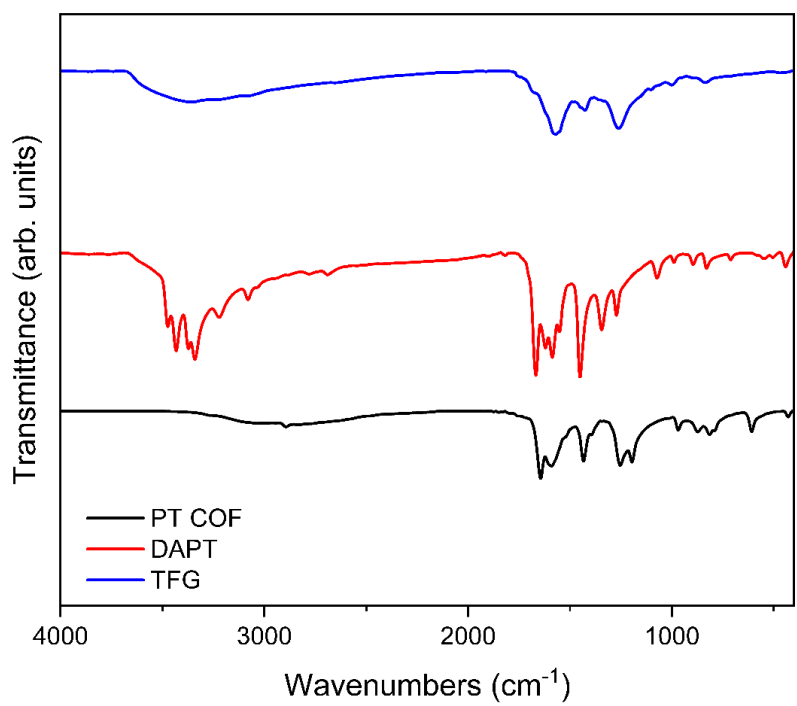

*Figure S19: FT-IR spectra of TFG, DAPT (precursors), and PT-COF.*

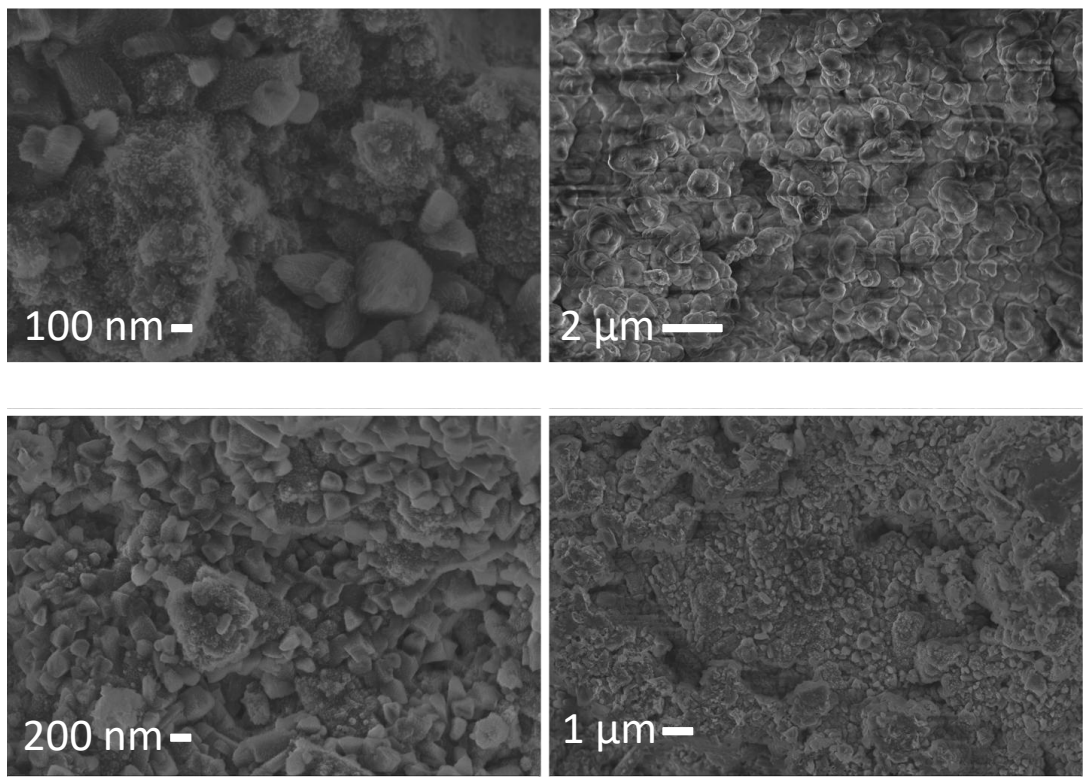

*Figure S20: Scanning electron microscopy (SEM) images of PT-COF.*

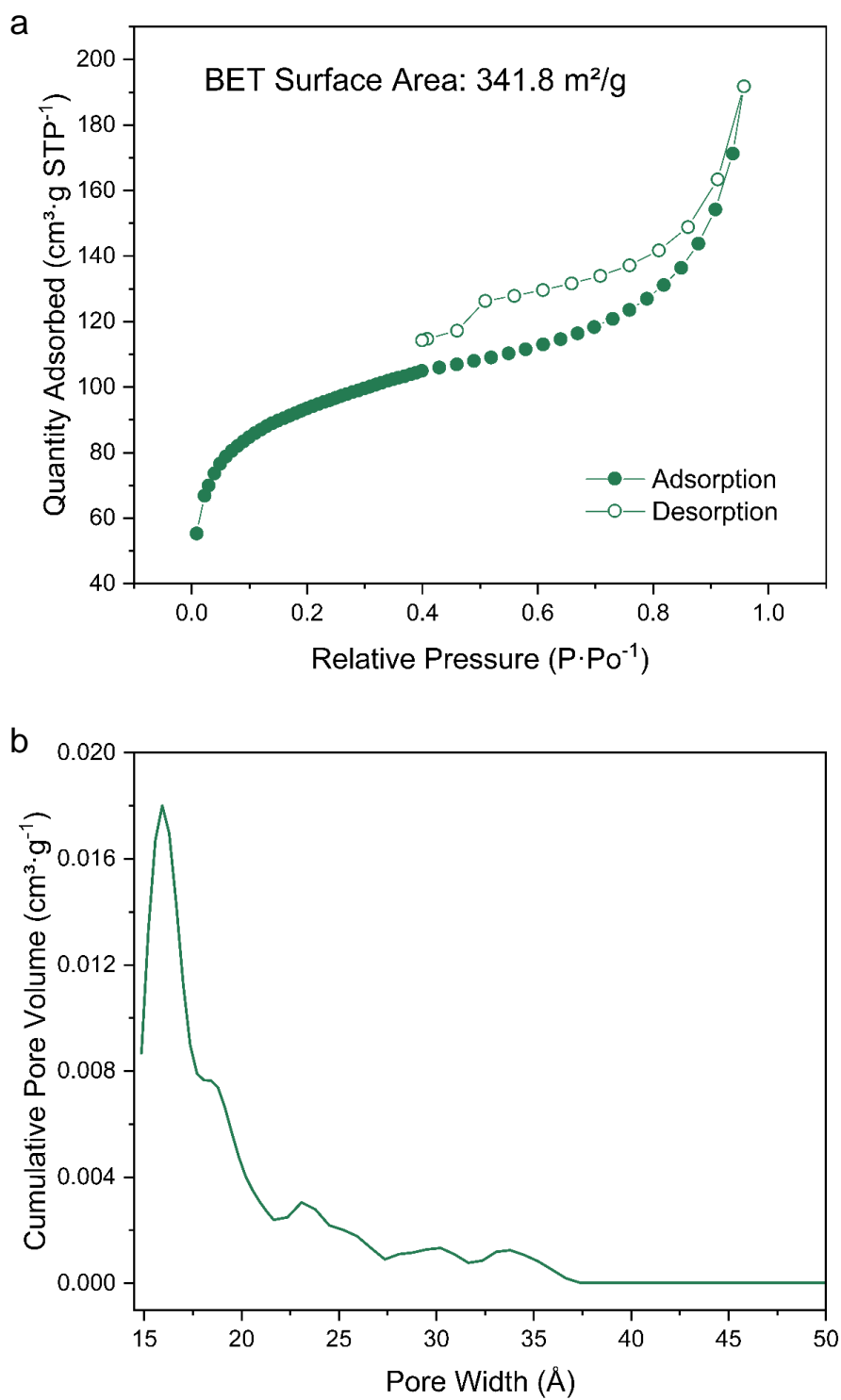

*Figure S21: a) Nitrogen adsorption isotherm of PT-COF. b) The pore size distribution of PT-COF exhibits a dominant peak centered at 1.6 nm.*

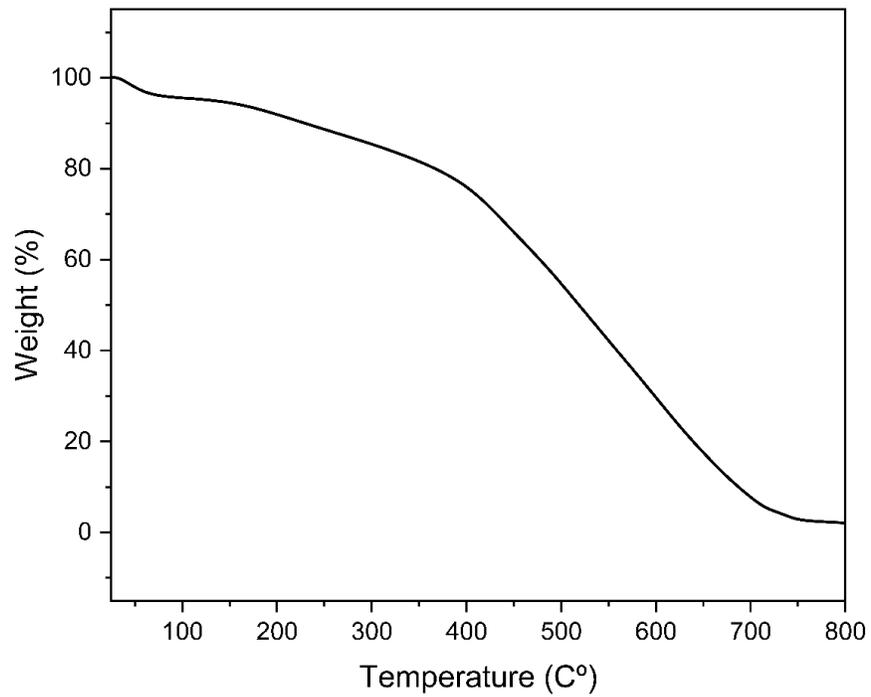

*Figure S22: Thermogravimetric analysis (TGA) of PT-COF recorded under nitrogen atmosphere up to 800 °C*

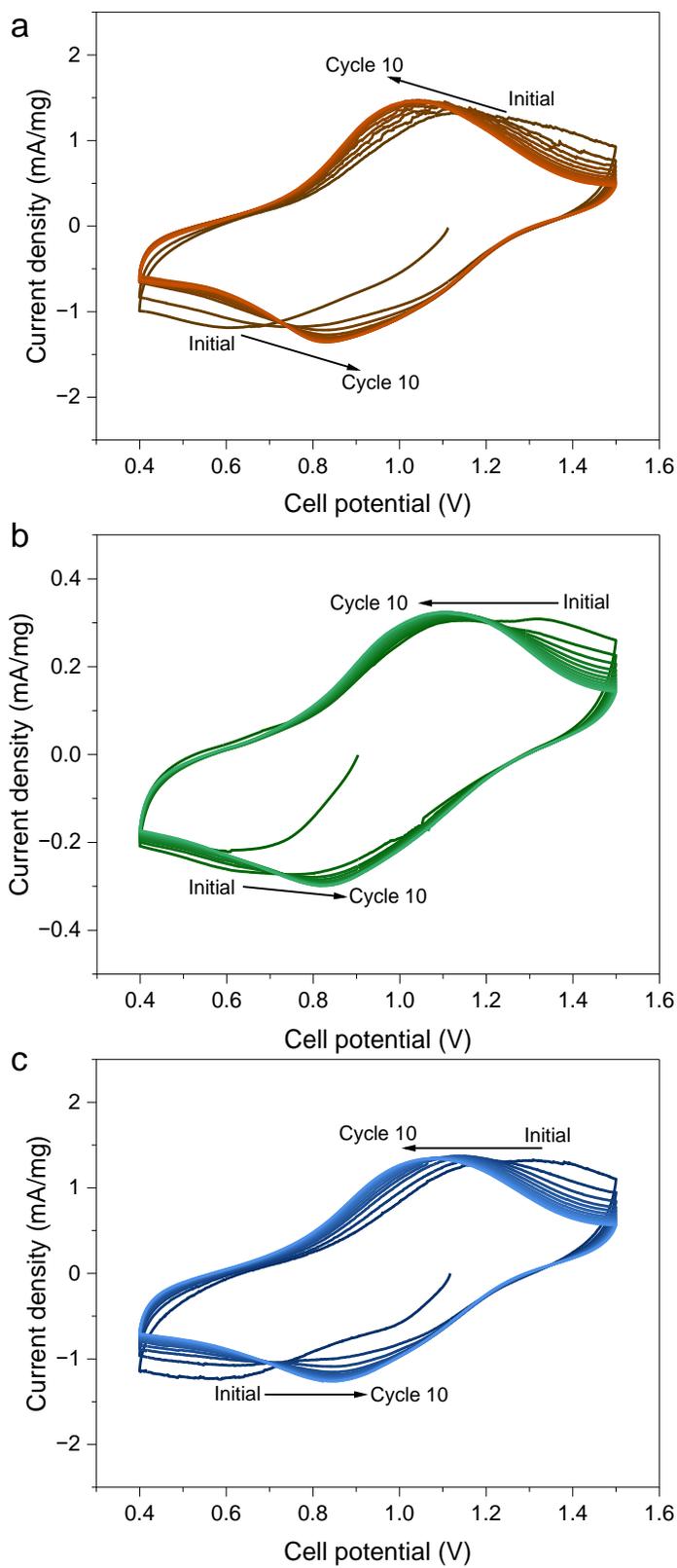

*Figure S23: CV curves utilising a COF-based cathode material for a) ZnAc, b) ZnAc20 and c) Zn**NH₄**Ac20.*

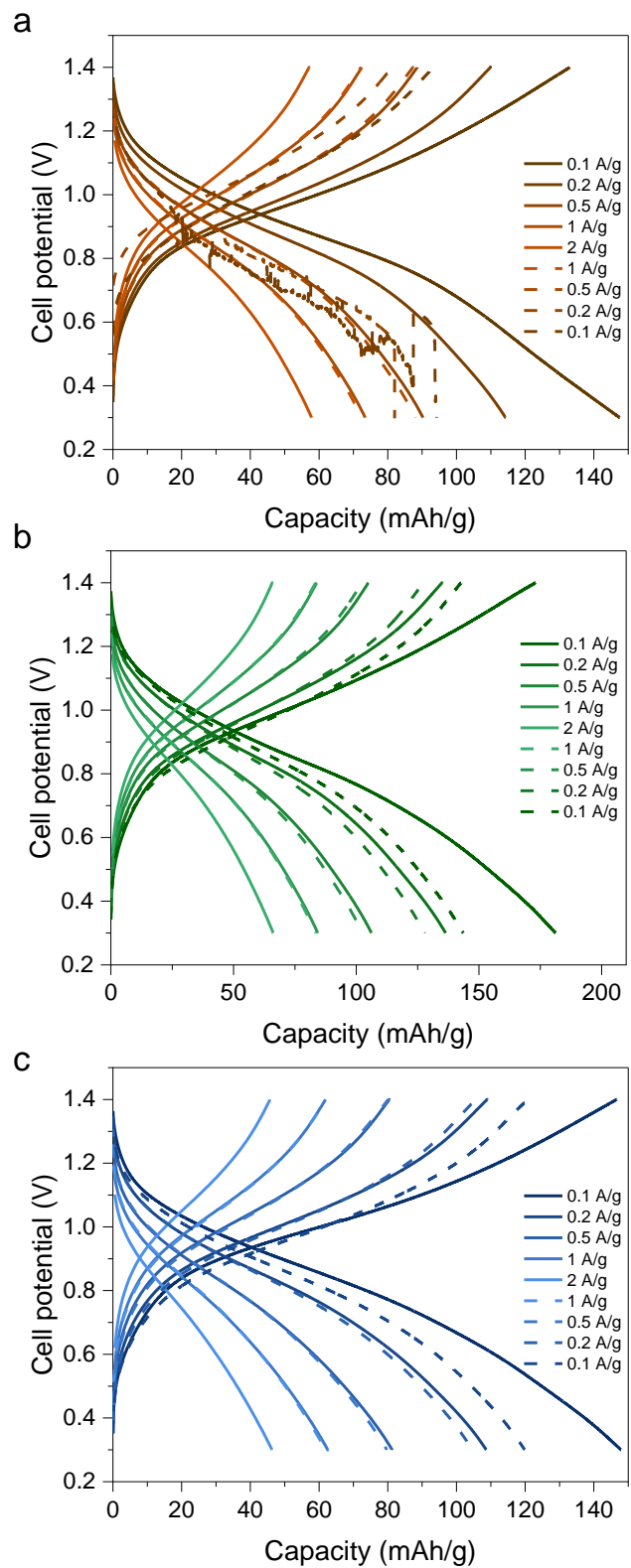

*Figure S24: GCD curves of the rate capability utilising COF-based cathode for a) ZnAc, b) ZnAc20 and c) ZnNH$_4$Ac20.*

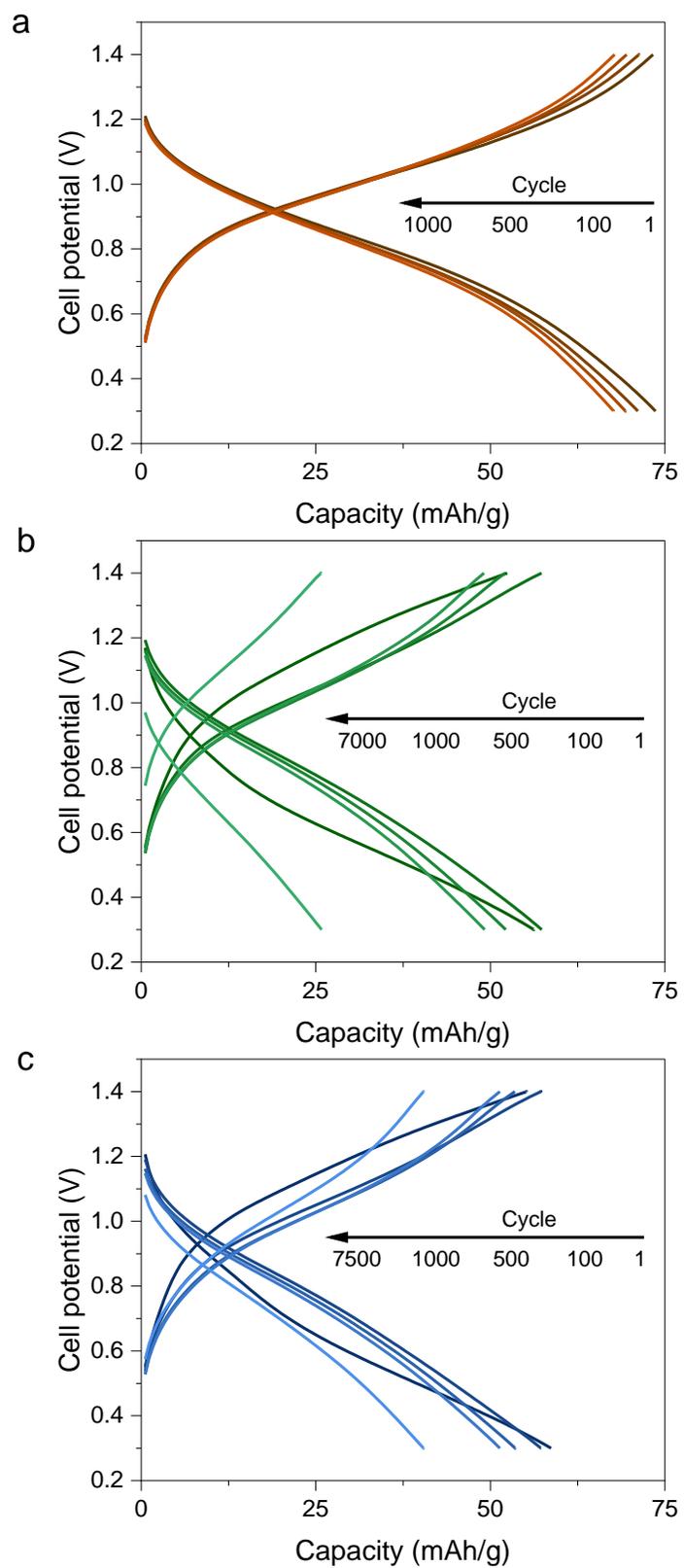

*Figure S25: GCD curves of the long-term stability test at 2 A/g utilising the COF-based cathode for a) ZnAc, b) ZnAc20 and c) Zn**NH₄**Ac20.*

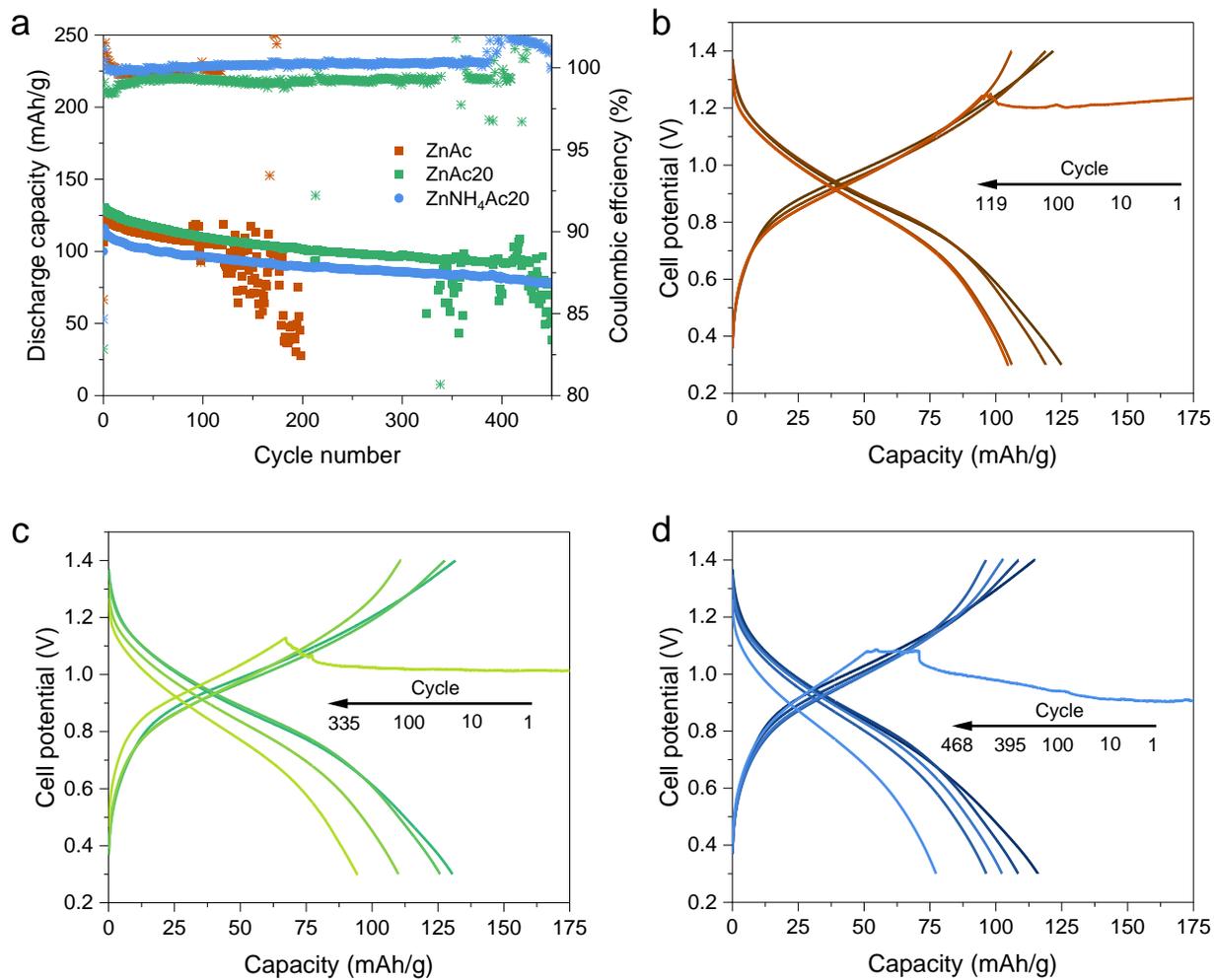

*Figure S26: Long-term GCD cycling of the COF cathode with 0.2 A/g. a) Comparison of the chosen electrolytes. GCD curve profiles for b) ZnAc, c) ZnAc20 and d) Zn**NH₄**Ac20.*

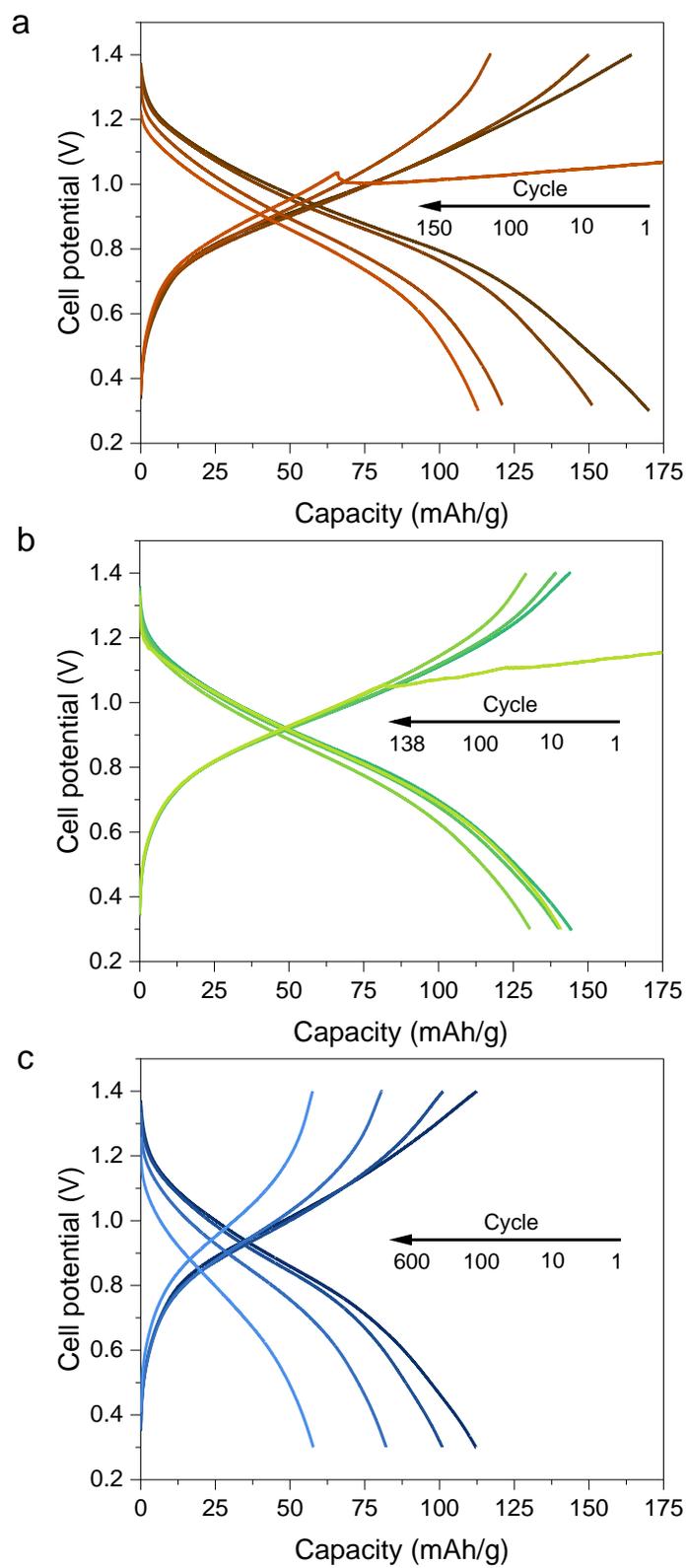

*Figure S27: Long-term GCD cycling of the COF cathode with 0.1 A/g for a) ZnAc, b) ZnAc20 and c) Zn**NH₄**Ac20.*

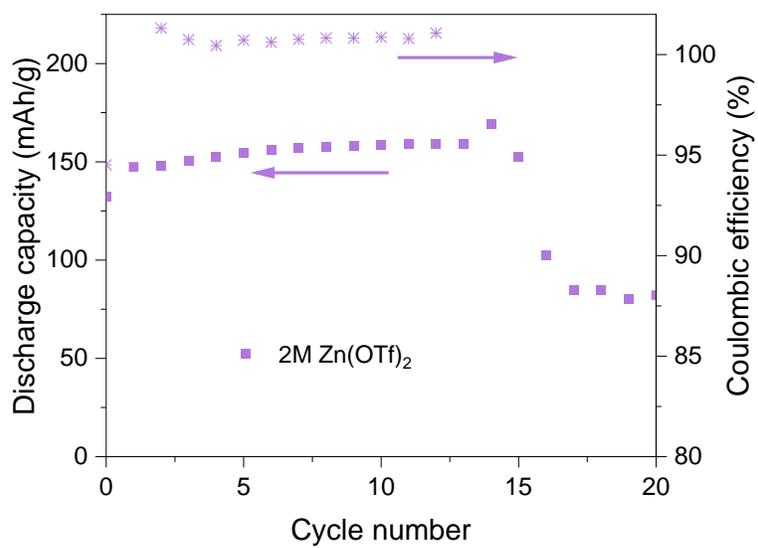

*Figure S28: Long-term GCD cycling of the COF cathode with 0.1 A/g for 2 M Zn(OTf)$_2$.*

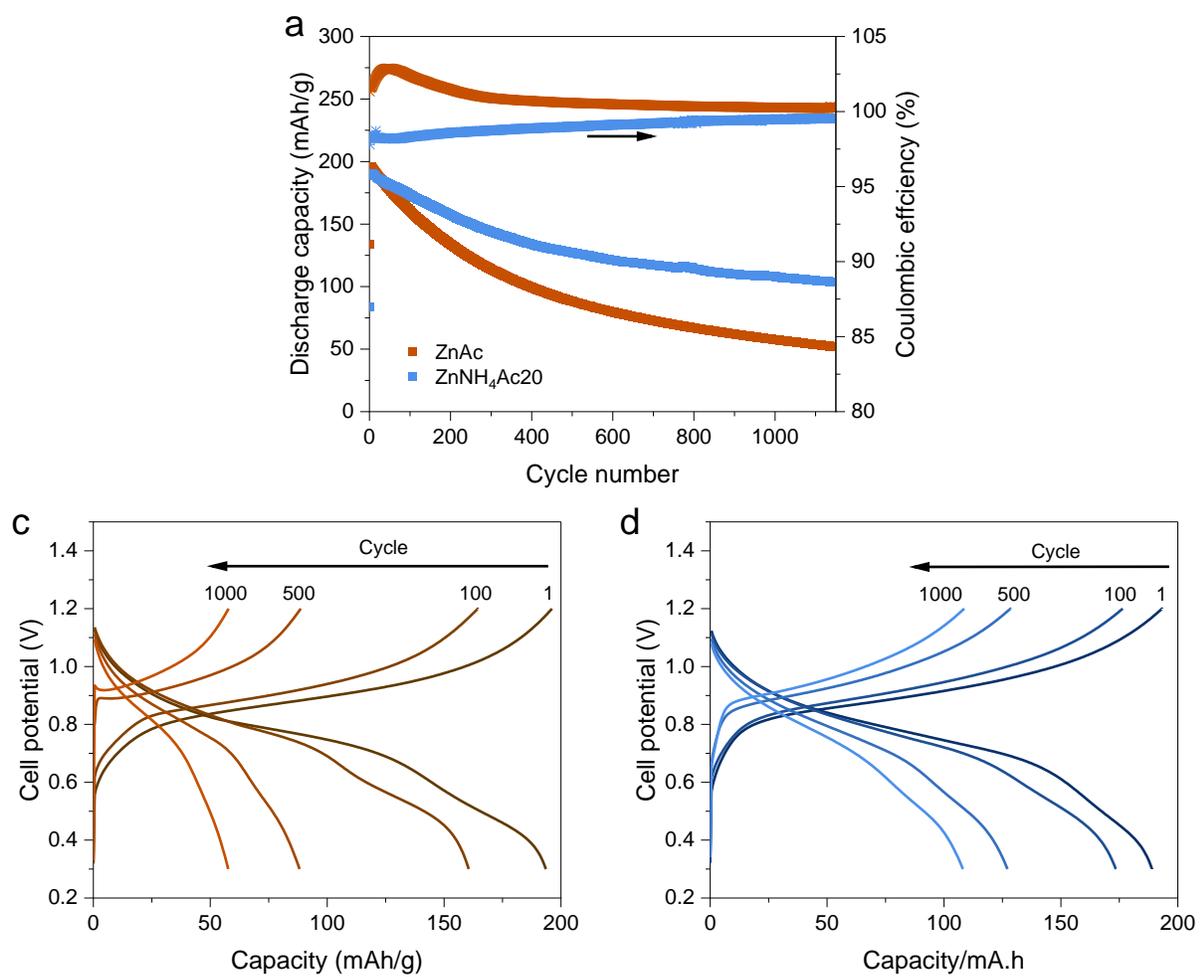

*Figure S29: Long-term GCD cycling of the NaVO cathode with 2 A/g. a) Comparison of the chosen electrolytes. GCD curve profiles for c) ZnAc and d) ZnAc20.*